# Low-energy trions in graphene quantum dots

H.-C. Cheng[1], N.-Y. Lue[2], Y.-C. Chen[1], G. Y. Wu[1,2]*

Department of Electrical Engineering
National Tsing-Hua University
Hsin-Chu, Taiwan 300

**Abstract**

We investigate, within the envelope function approximation, the low-energy states of trions in graphene quantum dots. The presence of valley pseudospin in graphene as an electron degree of freedom apart from spin adds convolution to the interplay between exchange symmetry and the electron-electron interaction in the trion, leading to new states of trions as well as a low energy trion spectrum different from those in semiconductors. Due to the involvement of valley pseudospin, it is found that the low-energy spectrum is nearly degenerate and consists of states all characterized by having an antisymmetric '(pseudospin)$\otimes$(spin)' component in the wave function, with the spin (pseudospin) part being either singlet (triplet) or triplet (singlet), as opposed to the spectrum in a semiconductor whose ground state is known to be nondegenerate and always a spin singlet in the case of $X^-$ trions. We investigate trions in the various regimes determined by the competition between quantum confinement and electron-electron interaction, both analytically and numerically. The numerical work is performed within a variational method accounting for electron mass discontinuity across the QD edge. The result for electron-hole correlation in the trion is presented. Effects of varying quantum dot size and confinement potential strength on the trion binding energy are discussed. The "relativistic effect" on the trion due to the unique relativistic type electron energy dispersion in graphene is also examined.

[1]Department of Physics, National Tsing-Hua University, Hsin-Chu, Taiwan 300
[2]Department of Electrical Engineering, National Tsing-Hua University, Hsin-Chu, Taiwan 300
*Corresponding author, email address: yswu@ee.nthu.edu.tw

I. Introduction

Graphene is a hexagonal, single layer of carbon atoms. It has risen as an important two-dimensional (2D) material for both scientific research and wide applications.[1] With electrons in graphene behaving as 2D Dirac fermions, it exhibits various novel phenomena, including Klein tunneling and unusual quantum Hall effect [1-3]. In addition, the graphene band structure shows unique two-fold valley degeneracy, thus bringing a novel electron degree of freedom (DOF) known as valley pseudospin which enriches physics as well as applications. Specifically, the presence of valley pseudospin leads to the valley-dependent physics[4-6] and the possibility of valley-based electronics (valleytronics)[4-10].

Although the free-standing monolayer graphene is gapless, it can also be made gapped, in principle, by growing it on a lattice matched *h*-BN substrate[11]. In either case, conduction band minima and valence band maxima occur at the Dirac points of Brillouin zone labeled K and K', with the band structure near each point given by the relativistic type dispersion - $E(\vec{k}) = \pm \hbar v_F |\vec{k}|$ ($v_F$ = Fermi velocity and $\vec{k}$ is defined relative to the Dirac point) in the gapless case or $E(\vec{k}) = \pm\sqrt{\Delta^2 + (\hbar v_F \vec{k})^2}$ ($2\Delta$ = band gap) in the gapped case. In addition to monolayer graphene, bilayer graphene can also be made gapped by applying a DC bias between the two graphene planes, with the resultant gap ($2\Delta$) being given by the DC voltage applied.[12,13] Since gapped graphene is a semiconductor, confined structures such as quantum wires or quantum dots (QDs) can be patterned in gapped graphene as in ordinary semiconductors, by just placing metal gates on graphene and applying gate voltages.[14-16] This provides nanoscale band gap-based confinement of carriers and makes graphene a promising material for nanoelectronic applications.

In this work, we consider carriers of opposite charges simultaneously confined in a QD in gapped graphene. Elementary few-particle states such as trions or excitons can be formed here just as in semiconductors. Being consisting of a small number of interacting particles, these states are relatively simple and reflect directly the effects of both quantum confinement and the inter-carrier interaction on the carriers. Therefore, we expect that a theoretical study of such states will aid future experiments which probe these few-particle states as a way to characterize graphene-based nanostructures as well as investigate the few-electron interaction physics in the structures. This is one of the reasons which motivate the present study. Additional reasons are stated below.

Among the few-particle systems, trions have been extensively studied in semiconductors both theoretically and experimentally.[17-21] A trion is a bound state consisting of three particles, e.g., two electrons and one hole in the case of a $X^-$ trion or one electron and two holes in the case of a $X^+$ trion, and provides an interesting example closely analogous to an ionized atom or molecule. For a trion with heavy holes, it mimics the ion $H^-$ in the case of an $X^-$ trion or $H_2^+$ in the case of an $X^+$ trion. Moreover, it exhibits pronounced interplay between the trion energy and the exchange symmetry (between the two electrons in an $X^-$ trion, for example), which is largely absent in excitons. For example, it is known that for an $X^-$ trion, a spin singlet is the ground state and is lower in energy than a triplet, due to the interplay.[19] Generally, there is also the interplay between the trion binding energy and the dimensionality. For example, the trion in a bulk semiconductor exhibits a 3D character with a small binding energy, while in CdTe-,[17-19] GaAs[17,19,20]- or ZnSe[19]- based quantum wells, it is quasi-2D and shows a significant binding energy, due to the enhanced electron-hole interaction in reduced dimension. Recently, the various interplays described above have been re-visited by Shiau et al.[21], based on an efficient theoretical method employing electron-exciton basis states for expanding the trion state, in the case of $X^-$ trions for example.

Graphene-based trions differ from semiconductor-based ones in several aspects. First and foremost, carriers in graphene carry both valley pseudospin and spin. Although the two types of spin are shown to be strongly analogous to each other – both carry magnetic moments in the case of gapped graphene, for example,[5,6,10] – the fact that they are independent DOFs is expected to come into play in the formation of trions, for instance, by adding interesting convolution to the interplay between the exchange symmetry and the trion energy.[22] Specifically, this work finds that the involvement of valley pseudospin results in new states of trions different from those in semiconductors, and a low energy trion state in graphene can be a spin singlet or triplet, as opposed to the spin singlet-only ground state in semiconductors.

The role of valley pseudospin as analyzed in the present work is summarized in the following. Let us assume that a K-valley, up-spin electron, denoted as $K\uparrow$, is initially present in the QD, and consider the optical excitation of an electron-hole pair in the QD, in the presence of $K\uparrow$. This excitation creates various types of low-energy $X^-$ trions, with the following electron parts, e.g.,$(K\uparrow,K\downarrow)$, $(K\uparrow,K'\uparrow)$, or $(K\uparrow,K'\downarrow)$ (all with the hole part being hidden). We focus on the usual limit where the characteristic length of the system (e.g., the QD radius or the exciton Bohr radius, whichever is shorter) is much greater than the lattice constant, and employ the envelope function approximation (EFA)[23,24] to write the two-electron part of the trion as follows,

$$F(\vec{r}_1,\vec{r}_2)\psi_K(\vec{r}_1)\psi_K(\vec{r}_2) \otimes (\text{spin singlet}),$$

$$[F(\vec{r}_1,\vec{r}_2)\psi^{(c)}{}_K(\vec{r}_1)\psi^{(c)}{}_{K'}(\vec{r}_2) - F(\vec{r}_2,\vec{r}_1)\psi^{(c)}{}_K(\vec{r}_2)\psi^{(c)}{}_{K'}(\vec{r}_1)] \otimes (\text{spin triplet}), \qquad (1)$$

$$[F(\vec{r}_1,\vec{r}_2)\psi^{(c)}{}_K(\vec{r}_1)\psi^{(c)}{}_{K'}(\vec{r}_2) + F(\vec{r}_2,\vec{r}_1)\psi^{(c)}{}_K(\vec{r}_2)\psi^{(c)}{}_{K'}(\vec{r}_1)] \otimes (\text{spin singlet}).$$

Here, we have ignored the spin-orbit interaction which is known to be rather weak in graphene,[3] and have therefore chosen the states to be the eigenstates of the total spin $S = s_1 + s_2$, which are either a singlet or triplet. All the states in Eqn. (1) obey the exchange symmetry required for fermions. The orbital part has been decomposed, within the EFA, into a slowly varying envelope function $F(\vec{r}_1,\vec{r}_2)$ and a fast varying $\psi^{(c)}{}_{K(K')}(\vec{r})$, where $\psi^{(c)}{}_{K(K')}(\vec{r})$ = conduction band Bloch function at the Dirac point K(K'). Depending on the symmetry of $F(\vec{r}_1,\vec{r}_2)$, Eqn. (1) leads to

$$F_S(\vec{r}_1,\vec{r}_2)\psi^{(c)}{}_K(\vec{r}_1)\psi^{(c)}{}_K(\vec{r}_2) \otimes (\text{spin singlet}),$$

$$F_S(\vec{r}_1,\vec{r}_2)[\psi^{(c)}{}_K(\vec{r}_1)\psi^{(c)}{}_{K'}(\vec{r}_2) - \psi^{(c)}{}_K(\vec{r}_2)\psi^{(c)}{}_{K'}(\vec{r}_1)] \otimes (\text{spin triplet}),$$

$$F_A(\vec{r}_1,\vec{r}_2)[\psi^{(c)}{}_K(\vec{r}_1)\psi^{(c)}{}_{K'}(\vec{r}_2) + \psi^{(c)}{}_K(\vec{r}_2)\psi^{(c)}{}_{K'}(\vec{r}_1)] \otimes (\text{spin triplet}), \qquad (2)$$

$$F_S(\vec{r}_1,\vec{r}_2)[\psi^{(c)}{}_K(\vec{r}_1)\psi^{(c)}{}_{K'}(\vec{r}_2) + \psi^{(c)}{}_K(\vec{r}_2)\psi^{(c)}{}_{K'}(\vec{r}_1)] \otimes (\text{spin singlet}),$$

$$F_A(\vec{r}_1,\vec{r}_2)[\psi^{(c)}{}_K(\vec{r}_1)\psi^{(c)}{}_{K'}(\vec{r}_2) - \psi^{(c)}{}_K(\vec{r}_2)\psi^{(c)}{}_{K'}(\vec{r}_1)] \otimes (\text{spin singlet}).$$

Here, $F_{S(A)}(\vec{r}_1,\vec{r}_2)$ refers to a symmetric (antisymmetric) envelope function with respect to the exchange between $\vec{r}_1$ and $\vec{r}_2$. In contrast, an X$^-$ trion in the semiconductor with nondegenerate conduction band minimum has the following two-electron part[19]

$$F_S(\vec{r}_1,\vec{r}_2) \otimes (\text{spin singlet}) \qquad (\text{for ground state trion}),$$

$F_A(\vec{r}_1,\vec{r}_2) \otimes$ (spin triplet)    (for excited state trion).    (2')

Comparing Eqns. (2) and (2'), we see clearly that with the electrons being in the same K valley, the 1$^{st}$ state in Eqn. (2) is a close analogy of the spin singlet state in Eqn. (2'), as if the valley degeneracy has been lifted. In contrast, other states in Eq. (2) contain combinations of products formed from opposite pseudospins (called pseudospin singlet / triplet), and represent new states of trions which have no analogue in Eqn. (2'). In general, if we compare the symmetry of wave functions in Eqns. (2) and (2'), it shows that there is a correspondence between the '*spin*' component in Eqn. (2') and the composite '*pseudospin* $\otimes$ *spin*' in Eqn. (2), as given in the following

**graphene-based trion**                                **semiconductor-based trion**

$\{\psi^{(c)}{}_K(\vec{r}_1)\psi^{(c)}{}_K(\vec{r}_2) \otimes (\text{spin singlet})\}$  $\rightarrow$    (spin singlet)

$\{[\psi^{(c)}{}_K(\vec{r}_1)\psi^{(c)}{}_{K'}(\vec{r}_2) - \psi^{(c)}{}_K(\vec{r}_2)\psi^{(c)}{}_{K'}(\vec{r}_1)] \otimes (\text{spin triplet})\}$ $\rightarrow$    (spin singlet)

$\{[\psi^{(c)}{}_K(\vec{r}_1)\psi^{(c)}{}_{K'}(\vec{r}_2) + \psi^{(c)}{}_K(\vec{r}_2)\psi^{(c)}{}_{K'}(\vec{r}_1)] \otimes (\text{spin singlet})\}$ $\rightarrow$    (spin singlet)

$\{[\psi^{(c)}{}_K(\vec{r}_1)\psi^{(c)}{}_{K'}(\vec{r}_2) + \psi^{(c)}{}_K(\vec{r}_2)\psi^{(c)}{}_{K'}(\vec{r}_1)] \otimes (\text{spin triplet})\}$ $\rightarrow$    (spin triplet)

$\{[\psi^{(c)}{}_K(\vec{r}_1)\psi^{(c)}{}_{K'}(\vec{r}_2) - \psi^{(c)}{}_K(\vec{r}_2)\psi^{(c)}{}_{K'}(\vec{r}_1)] \otimes (\text{spin singlet})\}$ $\rightarrow$    (spin triplet)

Thus, an antisymmetric (symmetric) *pseudospin* $\otimes$ *spin* in graphene plays the role of a spin singlet (triplet) in the semiconductor. Since the spin singlet (triplet) trion is the ground (excited) state in the semiconductor, we identify, in the graphene case, the states with antisymmetric (symmetric) *pseudospin* $\otimes$ *spin*, or correspondingly the states with symmetric $F_S(\vec{r}_1,\vec{r}_2)$ (antisymmetric $F_A(\vec{r}_1,\vec{r}_2)$), as the low energy (excited) trion states. For either the semiconductor- or graphene- based trions, the states with a symmetric $F_S(\vec{r}_1,\vec{r}_2)$ tend to place both of the electrons in the lowest orbital shell around the hole, giving a configuration similar to the ground state of H$^-$. This leads to the identification of these states as the low-energy states.

Apart from the involvement of valley pseudospin, there are other contrasts between the semiconductor-based trion and the trion considered by us here. First, a typical semiconductor-based trion involves a complicated valence band structure with different types of holes - heavy or light holes. However, in the graphene case, except for the presence of valley degeneracy, the band structure is relatively simple. It exhibits the electron-hole symmetry, as well as nondegenerate conduction and valence bands[3]. Second, we are primarily concerned here with the trion that is confined in a QD. On one hand, the presence of confinement destroys the translational symmetry and thus increases the level of theoretical difficulty. On the other hand, it gives us a chance to study trions in different regimes for a given QD. For example, denote the QD radius by 'R'. A variation in the gap parameter $\Delta$ (by applying a DC bias, for instance, in bilayer graphene) would bring about a variation in the carrier effective mass $m^*$ ($= \Delta/v_F^2$) and correspondingly also a variation in the exciton Bohr radius $a_B^*$ ($= O(\hbar^2 \varepsilon / m^* e^2)$, $\varepsilon$ = effective dielectric constant). This leads to different regimes for the given QD, depending on the competition between the QD confinement energy $E_{QD}$ ($= O(h^2/m^* R^2)$) and the e-e interaction energy ($= \max(E_{e-e}^{(C)}, E_{ex})$). Here, $E_{e-e}^{(C)} = O(e^2/\varepsilon R)$ being the QD charging energy and $E_{ex} = O(e^2/\varepsilon a_B^*)$ being the exciton binding energy). For $R \ll a_B^*$, the QD confinement dominates, while for $R \gg a_B^*$, the interaction dominates. The free trion typically studied in semiconductors is a well known example which belongs to the strong interaction regime. Last, semiconductor trions consist of carriers which are usually taken to have parabolic energy dispersions and constant effective masses. However, in the graphene case, due to the unique relativistic type energy dispersion, high energy carriers have a "relativistic", energy (or momentum) dependent effective mass. Therefore, depending on the magnitude of energy of constituent carriers involved, interesting "relativistic effects" may emerge in graphene trions.

It is also noted that the study of trions in a gapped graphene-based QD constitutes a significant step towards eventually developing graphene-based long distance quantum communications. In a previous study,[10] it was shown that a graphene qubit (comprising of a pair of coupled QDs in gapped graphene, with two electrons separately confined in the QDs) based on valley pseudospin can serve as a quantum memory in the communication. Such utilization involves the interaction between a photon and one of the QDs. When the photon is absorbed, it creates an electron-hole pair, and a possible formation of $X^-$ trion in the QD.

Our work studies the low-energy states of a trion in a type-I gapped graphene-based QD, where electrons and holes are both confined in the QD. For simplicity, we consider monolayer graphene on $h$-BN as the prototype system in view of its simple electron energy dispersion $E(\vec{k}) = \pm\sqrt{\Delta^2 + (\hbar v_F \vec{k})^2}$, but the discussion can be generalized to bilayer graphene as well.[25] Due to the electron-hole symmetry between X$^-$ and X$^+$, we shall focus only on X$^-$. The presentation is organized as follows. In **Sec. II**, we present the theoretical formulation for X$^-$ trions involving valley pseudospin, in the envelope function approximation. In **Sec. III**, we discuss the low-energy trion spectrum as well as trions in different regimes. In **Sec. IV**, we describe the variational method for the calculation of low-energy trion states. In **Sec. V**, numerical results of binding energies as well as wave functions of trions are presented. In **Sec. VI**, we summarize the work. **Appendix A** lists the various parameters of basis functions used in the variational calculation. **Appendix B** presents the exact one-carrier ground state solution in the QD, which can be used to provide the ground state energy of the noninteracting system of two electrons and one hole, as reference energy for the calculation of trion binding energy. **Appendix C** provides useful formulas for the various integrals involved in the variational calculation, such as the overlap or the Hamiltonian matrix element between two basis functions. **Appendix D** discusses the approximation involved in our treatment and provides an assessment of the numerical error in the calculation.

## II.     The theoretical formulation for X$^-$ trions involving valley pseudospin

We present the theoretical formulation for X$^-$ trions in gapped graphene (band gap = 2Δ), in the envelope function approximation. In the case where the constituent carriers in the trion are near the conduction and valence band edges, we are in the nonrelativistic type regime (also called Schrodinger regime in this work), where the carrier energy with respect to the band edge is given by $E_c(k) \approx \frac{\hbar^2 k^2}{2m^*}$ for an electron or $E_v(k) \approx -\frac{\hbar^2 k^2}{2m^*}$ for a hole. Here, the effective mass m$^*$ = $\Delta/v_F^2$, and $k$ = wave vector relative to K or K'. In the case where the constituent carriers are away from the band edges, we are in the relativistic type regime (also called the Dirac regime in this work), where the effective mass becomes k-dependent and is given by $m^*(k) = \frac{1}{2}\left[\left(\hbar^2 k^2 / v_F^2 + \Delta^2 / v_F^4\right)^{1/2} + \Delta/v_F^2\right]$. Both regimes will be considered. We focus on the specific example where the X$^-$ trion consists of $(K\uparrow, K'\uparrow, \overline{K}'\downarrow)$. $\overline{K}'$ here denotes a K' hole. The discussion below can be generalized to other cases.

### The field operator involving the valley pseudospin

We employ the 2$^{nd}$ quantization formalism to present the theoretical formulation. Let the one-particle Hilbert space consist of both conduction and valence band states. The corresponding annihilation field operator is written as

$$\Psi(r) \approx \Psi_{K\uparrow}(r) + \Psi_{K'\uparrow}(r) + \Psi_{\bar{K}'\downarrow}^{(+)}(r) + \Psi_{irrel.}(r),$$

$$\Psi_{K\uparrow}(r) \equiv \sum_{k\sim 0} K_{k\uparrow} \frac{\exp(ik \cdot r)}{\sqrt{\Omega}} \psi^{(c)}_{K}(r) \otimes \uparrow,$$

$$\Psi_{K'\uparrow}(r) \equiv \sum_{k\sim 0} K'_{k\uparrow} \frac{\exp(ik \cdot r)}{\sqrt{\Omega}} \psi^{(c)}_{K'}(r) \otimes \uparrow, \qquad (3)$$

$$\Psi_{\bar{K}'\downarrow}^{(+)}(r) \equiv \sum_{k\sim 0} \bar{K}'^{(+)}_{k\downarrow} \frac{\exp(-ik \cdot r)}{\sqrt{\Omega}} \psi^{(v)}_{K'}(r) \otimes \uparrow.$$

Here, $k$ = wave vector relative to K or K', and $\Omega$ = total system area. $K_{k\uparrow}$ ($K'_{k\uparrow}$) removes a $K\uparrow$ ($K'\uparrow$) electron at $k$ in the conduction band, and $\bar{K}'^{(+)}_{k\downarrow}$ creates a $K'\downarrow$ hole at k (or removes a $K'\uparrow$ electron at -k) in the valence band. $\psi^{(c)}_{K(K')}(r)$ ($\psi^{(v)}_{K(K')}(r)$) denotes the conduction (valence) band Bloch wave function at K (K'). Here, we have taken the normalization $\int_{cell} |\psi^{(c)}_{K(K')}(\vec{r})|^2 d^2r = \int_{cell} |\psi^{(v)}_{K(K')}(\vec{r})|^2 d^2r = \Omega_{cell}$ ($\Omega_{cell}$ = area of a unit cell). $\Psi_{irrel.}(r)$ represents the part of $\Psi(r)$ irrelevant for the discussion of the specific type of trion considered here, for example, which removes a $K\downarrow$ electron in the conduction band.

We note that Eqn. (3) employs the approximation where the exact wave function at k is expressed in terms of that at k = 0, e.g., $\psi^{(c)}_{K(K')}(r)$ or $\psi^{(v)}_{K(K')}(r)$. For trions consisting of the low-energy states near Dirac points, this approximation works well. **Appendix D** provides an analysis concerning the validity of this approximation in the present study.

### The envelope function and the trion wave function

Next, we describe the trion state. Let $|0\rangle$ = ground state of intrinsic gapped graphene, which has a filled valence band and an empty conduction band. In terms of $|0\rangle$, we write the trion state as

$$|\psi_{trion}\rangle = \sum_{k_1 k_2 k_3} f_{k_1 k_2 k_3} |k_1 k_2 k_3\rangle,$$
$$|k_1 k_2 k_3\rangle \equiv K^+_{k_1 \uparrow} K'^+_{k_2 \uparrow} \bar{K}'^+_{k_3 \downarrow} |0\rangle. \qquad (4)$$

$\{|k_1 k_2 k_3\rangle\text{'s}\}$ represent the states of two electrons and one hole, all being noninteracting, and form the set of basis functions for expanding the trion state. $\{f_{k_1 k_2 k_3}\text{'s}\}$ are the corresponding expansion coefficients subject to the normalization $\sum_{k_1 k_2 k_3} |f_{k_1 k_2 k_3}|^2 = 1$. Eqn. (4) gives a description of the trion state in the k-space, with $f_{k_1 k_2 k_3}$ being the k-space wave function. $\{f_{k_1 k_2 k_3}\text{'s}\}$ can be transformed to the r-space, giving the envelope function

$$F_{trion}(r_1, r_2, r_3) = \sum_{k_1 k_2 k_3} f_{k_1 k_2 k_3} \frac{\exp(ik_1 \cdot r_1)}{\sqrt{\Omega}} \frac{\exp(ik_2 \cdot r_2)}{\sqrt{\Omega}} \frac{\exp(-ik_3 \cdot r_3)}{\sqrt{\Omega}} \qquad (5)$$

subject to the normalization $\int |F_{trion}(\vec{r}_1, \vec{r}_2, \vec{r}_3)|^2 d^2\vec{r}_1 d^2\vec{r}_2 d^2\vec{r}_3 = 1$. $F_{trion}(r_1, r_2, r_3)$ is closely related to the following wave function, $\Psi_{trion}(r_1, r_2, r_3)$, defined in the r-space by

$$\Psi_{trion}(r_1, r_2, r_3)$$
$$\equiv \frac{1}{\sqrt{2}} \langle 0 | \Psi_{\bar{K}\downarrow}(r_3) [\Psi_{K\uparrow}(r_2) + \Psi_{K'\uparrow}(r_2)][\Psi_{K\uparrow}(r_1) + \Psi_{K'\uparrow}(r_1)] |\psi_{trion}\rangle \qquad (6)$$

which gives the coordinate representation of $|\psi_{trion}\rangle$. Note that $\Psi_{trion}(r_1, r_2, r_3)$ given by Eqn. (6) is antisymmetric with respect to the exchange between $r_1$ and $r_2$. Substituting the various field operators introduced in Eqn. (3) into the above equation, we express $\Psi_{trion}(r_1, r_2, r_3)$, in terms of $F_{trion}(r_1, r_2, r_3)$, as

$$\Psi_{trion}(r_1, r_2, r_3)$$
$$= \frac{1}{\sqrt{2}} \begin{bmatrix} F_{trion}(\vec{r}_1, \vec{r}_2, \vec{r}_3) \psi^{(c)}{}_K(\vec{r}_1) \psi^{(c)}{}_{K'}(\vec{r}_2) \\ - F_{trion}(\vec{r}_2, \vec{r}_1, \vec{r}_3) \psi^{(c)}{}_K(\vec{r}_2) \psi^{(c)}{}_{K'}(\vec{r}_1) \end{bmatrix} \otimes (\uparrow_1 \uparrow_2) \quad (7)$$
$$\left[ \psi^{(v)}{}_{K'}(\vec{r}_3) \right]^* \otimes (\uparrow_3)$$

$\Psi_{trion}(r_1, r_2, r_3)$ can be interpreted as the total wave function of the trion, which includes both the envelope function and the Bloch wave function and, hence, describes the trion state down to the sub-cell details. In comparison, the envelope function $F_{trion}(r_1, r_2, r_3)$ is a coarse-grain average of $\Psi_{trion}(r_1, r_2, r_3)$ and describes the trion only on the length scale above the unit cell.

### **The trion Hamiltonian and the envelope function equation**

Now, we discuss the trion Hamiltonian $H_{trion}$ and the corresponding effective wave equation satisfied by $F_{trion}(r_1, r_2, r_3)$,

$$H_{trion} F_{trion}(r_1, r_2, r_3) = E F_{trion}(r_1, r_2, r_3). \quad (8)$$

This defines $H_{trion}$ as the effective Hamiltonian for the trion. Eqn. (8) (or $H_{trion}$) is derived as follows.

We start with the description of the many-electron Hamiltonian for the system, which is given by

$$H = \int d^2 r \Psi^+(r) H_{crystal} \Psi(r) + \frac{1}{2} \iint d^2 r_1 d^2 r_2 \Psi^+(r_1) \Psi^+(r_2) V_{e-e}(r_1 - r_2) \Psi(r_2) \Psi(r_1),$$
$$V_{e-e}(\vec{r}_1 - \vec{r}_2) = \frac{e^2}{4\pi\varepsilon |\vec{r}_1 - \vec{r}_2|}, \quad (9)$$

where $H_{crystal}$ denotes the one-electron crystal Hamiltonian for gapped graphene, with $H_{crystal} = -\frac{\hbar^2}{2m} \nabla^2 + V_{crystal}(\vec{r})$, $V_{e-e}$ denotes the electron-electron (e-e) interaction, and ε is an effective dielectric constant determined by graphene as well as the substrate supporting graphene.

The trion state satisfies the following Hamiltonian equation

$$H|\psi_{trion}\rangle = E|\psi_{trion}\rangle. \qquad (10)$$

Substituting the field operator $\Psi(r)$ given by Eqn. (3) into Eqn. (10), we obtain

$$\left\{ \begin{array}{l} \sum_k \left[ E_c(k)\left(K_{k\uparrow}^{(+)}K_{k\uparrow} + K'^{(+)}_{k\uparrow}K'_{k\uparrow}\right) - E_v(-k)\bar{K}'^{(+)}_{k\downarrow}\bar{K}'_{k\downarrow} \right] \\ + \sum_{q,k_1,k_2} \bar{V}_{e-e}(q) K_{k_1+q\uparrow}^{(+)} K'^{(+)}_{k_2-q\uparrow} K'_{k_2\uparrow} K_{k_1\uparrow} \\ - \sum_{q,k_1,k_3} \bar{V}_{e-e}(q) K_{k_1+q\uparrow}^{(+)} \bar{K}'^{(+)}_{k_3\downarrow} \bar{K}'_{k_3+q\downarrow} K_{k_1\uparrow} \\ - \sum_{q,k_2,k_3} \bar{V}_{e-e}(q) K'^{(+)}_{k_2+q\uparrow} \bar{K}'^{(+)}_{k_3\downarrow} \bar{K}'_{k_3+q\downarrow} K'_{k_2\uparrow} + H_{ignored} \end{array} \right\} |\psi_{trion}\rangle \qquad (11)$$

$$= E|\psi_{trion}\rangle$$

$E_c(k)$ and $E_v(k)$ are the conduction and valence band dispersions, respectively, and satisfy the following Hamiltonian equations

$$H_{crystal}\left[ e^{ik\cdot r}\psi^{(c)}_{K(K')}(r) \right] \approx E_c(k)\left[ e^{ik\cdot r}\psi^{(c)}_{K(K')}(r) \right],$$
$$H_{crystal}\left[ e^{ik\cdot r}\psi^{(v)}_{K(K')}(r) \right] \approx E_v(k)\left[ e^{ik\cdot r}\psi^{(v)}_{K(K')}(r) \right].$$

Here, $\bar{V}_{e-e}(q)$ is approximately the Fourier transform of $V_{e-e}(\vec{r})$, given by $\bar{V}_{e-e}(q) \approx \frac{1}{\Omega}\int d^2 r V_{e-e}(\vec{r})\exp(-i\vec{q}\cdot\vec{r})$. The last three lines in Eqn. (11) involving $\bar{V}_{e-e}(q)$ represent, respectively, the Coulomb interaction between a K↑ and a K'↑ electrons, between a K↑ electron and a K'↓ hole, and between a K'↑ electron and a K'↓ hole. $H_{ignored}$ here consists of terms typically ignored for the treatment of trions in semiconductors, for example, the spin-flipping electron-hole (e-h) exchange scattering, as well as additional terms only present in graphene due to the presence of valley pseudospin (see below).

To proceed further in Eqn. (11), we expand $|\psi_{trion}>$ in terms of $\{|k_1k_2k_3>$'s$\}$ according to Eqn. (4), and project each side of Eqn. (11) onto $|k_1k_2k_3>$. Dropping $H_{ignored}$, we obtain

$$\begin{aligned}Ef_{k_1k_2k_3} &= \sum_{k_1'k_2'k_3'}<k_1k_2k_3|H|k_1'k_2'k_3'>f_{k_1'k_2'k_3'}\\ &\approx [E_c(k_1)+E_c(k_2)-E_v(-k_3)]f_{k_1k_2k_3} + \sum_{k_1'k_2'}\overline{V}_{e-e}(k_1-k_1')f_{k_1'k_2'k_3}\\ &- \sum_{k_1'k_3'}\overline{V}_{e-e}(k_1-k_1')f_{k_1'k_2k_3'} - \sum_{k_2'k_3'}\overline{V}_{e-e}(k_2-k_2')f_{k_1k_2'k_3'}.\end{aligned} \quad (12)$$

$\{f_{k_1k_2k_3}$'s$\}$ above are coupled together by the Coulomb interaction between electrons as well as between electrons and holes. Multiplying Eqn. (12) with $\frac{\exp(ik_1\cdot r_1)}{\sqrt{\Omega}}\frac{\exp(ik_2\cdot r_2)}{\sqrt{\Omega}}\frac{\exp(-ik_3\cdot r_3)}{\sqrt{\Omega}}$ on each side and summing it over $k_1$, $k_2$, and $k_3$, we finally arrive at Eqn. (8), with

$$\begin{aligned}H_{trion} &\approx E_c(k_1\to -i\nabla_1)+E_c(k_2\to -i\nabla_2)-E_v(k_3\to -i\nabla_3)\\ &+\frac{e^2}{4\pi\varepsilon|\vec{r}_1-\vec{r}_2|}-\frac{e^2}{4\pi\varepsilon|\vec{r}_1-\vec{r}_3|}-\frac{e^2}{4\pi\varepsilon|\vec{r}_2-\vec{r}_3|}\\ &+V_e(\vec{r}_1)+V_e(\vec{r}_2)+V_h(\vec{r}_3).\end{aligned}$$

or more explicitly,

$$\begin{aligned}H_{trion} &\approx -\frac{\hbar^2}{2}\nabla_1\cdot\left(\frac{1}{m^*(\vec{r}_1)}\nabla_1\right)-\frac{\hbar^2}{2}\nabla_2\cdot\left(\frac{1}{m^*(\vec{r}_2)}\nabla_2\right)-\frac{\hbar^2}{2}\nabla_3\cdot\left(\frac{1}{m^*(\vec{r}_3)}\nabla_3\right)\\ &+\frac{e^2}{4\pi\varepsilon|\vec{r}_1-\vec{r}_2|}-\frac{e^2}{4\pi\varepsilon|\vec{r}_1-\vec{r}_3|}-\frac{e^2}{4\pi\varepsilon|\vec{r}_2-\vec{r}_3|}\\ &+V_e(\vec{r}_1)+V_e(\vec{r}_2)+V_h(\vec{r}_3).\end{aligned} \quad (13)$$

Here, we have generalized $H_{trion}$ to account for the variation of energy gap ($\Delta(\vec{r})$)

across the boundary of the QD. The gap variation provides, in $H_{trion}$, the quantum confinement potentials for electrons and holes, $V_{e(h)}(\vec{r})$. $V_{e(h)}(\vec{r})$ and the gap function $\Delta(\vec{r})$ satisfy the following identity

$$\Delta(r > R) - \Delta(r < R) = \frac{1}{2}[V_e(r > R) + V_h(|r > R) - V_e(r < R) - V_h(r < R)].$$

Fig. 1 shows $V_{e(h)}(\vec{r})$ in the type-I QD. Moreover, due to the gap variation, there is a corresponding mass variation in the space. The Hamiltonian in Eqn. (13) accounts for the variation by replacing the standard kinetic energy operator '$-\frac{\hbar^2}{2m^*}\nabla_i^2$' with

'$-\frac{\hbar^2}{2}\nabla_i \cdot \left(\frac{1}{m^*(r_i)}\nabla_i\right)$', where $m^*(r_i) = \Delta(r_i)/v_F^2$ in the Schrodinger regime and

$m^*(r_i) = \frac{1}{2}\left[\left(-\hbar^2\nabla_i^2/v_F^2 + \Delta(r_i)^2/v_F^4\right)^{1/2} + \Delta(r_i)/v_F^2\right]$ in the Dirac regime, for i = 1-3.

This replacement ensures the positive definiteness of the operator, and has often been used in the theory of semiconductor heterostructures.[24, 26]

Eqns. (8) and (13) together form the EFA description of trion states. This description can be improved by treating $H_{ignored}$ within the perturbation theory as a correction to the EFA. An example will be given in **Sec. III**.

### *H<sub>ignored</sub>*

Now, we describe $H_{ignored}$. First, we note that, although the expression of $H_{trion}$ given by Eqn. (13) looks exactly the same as the Hamiltonian commonly used for semiconductor-based trions, as if the valley pseudospin did not play any role, the approximations invoked in the derivation of the two Hamiltonians are actually different. In both cases, the approximation neglects higher-order terms such as the spin-flipping e-h exchange scattering given by

$$H_{e-h}^{(\uparrow\leftrightarrow\downarrow)} = \sum_{q,k_2,k_3} \bar{V}_{e-h}^{(\uparrow\leftrightarrow\downarrow)}(q) K'^{(+)}_{k_1-q\downarrow} \bar{K}'^{(+)}_{-k_1\uparrow} \bar{K}'_{-k_2-q\downarrow} K'_{k_2\uparrow}, \qquad (14\text{-a})$$

$$\overline{V}_{e-h}^{(\uparrow \leftrightarrow \downarrow)}(q)$$
$$\equiv \frac{1}{\Omega^2}\int d^2r_1 d^2r_2\, e^{-i\vec{q}\cdot(\vec{r}_2-\vec{r}_1)}\psi_{K'}^{(c)*}(\vec{r}_1)\psi_{K'}^{(v)*}(\vec{r}_2)V_{e-e}(\vec{r}_1-\vec{r}_2)\psi_K^{(v)}(\vec{r}_1)\psi_K^{(c)}(\vec{r}_2). \quad (14\text{-b})$$

Such a scattering swaps the spins of the electron and the hole in the trion.

In the case of graphene, $H_{ignored}$ further includes a term analogous to $H_{e-h}^{(\uparrow \leftrightarrow \downarrow)}$ – the valley-flipping e-h exchange scattering given by

$$H_{e-h}^{(K \leftrightarrow K')} = \sum_{q,k_2,k_3} \overline{V}_{e-h}^{(K \leftrightarrow K')}(q) K_{k_1-q\uparrow}^{(+)} \overline{K}_{-k_1\downarrow}^{(+)} \overline{K}'_{-k_2-q\downarrow} K'_{k_2\uparrow}, \quad (15\text{-a})$$

$$\overline{V}_{e-h}^{(K \leftrightarrow K')}(q)$$
$$\equiv \frac{1}{\Omega^2}\int d^2r_1 d^2r_2\, e^{-i\vec{q}\cdot(\vec{r}_2-\vec{r}_1)}\psi_{K'}^{(c)*}(\vec{r}_1)\psi_{K'}^{(v)*}(\vec{r}_2)V_{e-e}(\vec{r}_1-\vec{r}_2)\psi_K^{(v)}(\vec{r}_1)\psi_K^{(c)}(\vec{r}_2). \quad (15\text{-b})$$

This scattering annihilates an e-h pair in the K' valley and creates a new pair in the K valley. Such a scattering is absent in a typical semiconductor. Being analogous to the spin-flipping e-h exchange scattering, the above scattering is also neglected in the present EFA description.

Moreover, $H_{ignored}$ also includes the following valley-flipping e-e exchange scattering

$$H_{e-e}^{(K \leftrightarrow K')} = -\sum_{q,k_1,k_2} \overline{V}_{e-e}^{(K \leftrightarrow K')}(q) K_{k_2-q\uparrow}^{(+)} K'^{(+)}_{k_1+q\uparrow} K'_{k_2\uparrow} K_{k_1\uparrow}, \quad (16\text{-a})$$

$$\overline{V}_{e-e}^{(K \leftrightarrow K')}(q) \equiv$$
$$\frac{1}{\Omega^2}\int d^2r_1 d^2r_2\, e^{-i\vec{q}\cdot(\vec{r}_2-\vec{r}_1)}\psi_{K'}^{(c)*}(\vec{r}_1)\psi_{K'}^{(c)*}(\vec{r}_2)V_{e-e}(\vec{r}_1-\vec{r}_2)\psi_{K'}^{(c)}(\vec{r}_1)\psi_K^{(c)}(\vec{r}_2), \quad (16\text{-b})$$

which is also absent in the semiconductor with a nondegenerate conduction band minimum. The matrix element $\overline{V}_{e-e}^{(K \leftrightarrow K')}(q)$ given above for the scattering indicates that a large intervalley momentum transfer is involved in the scattering. In comparison to the non-valley-flipping e-e scattering already included in $H_{trion}$, we regard the present scattering as a higher-order effect and place it in $H_{ignored}$.

### **Definition of trion binding energy**

Before we close this section, we introduce the definition of trion binding energy as follows,

$$E_{binding} = E(e = 0) - E,$$

where '$E$' is the eigenstate energy in Eqn. (8). '$E(e = 0)$' is the ground state energy of two electrons and one hole, all being noninteracting with the charge (e) being turned off. Obviously $E(e = 0)$ is determined by the ground state energy of each carrier, e.g., $E(e = 0) = 2E_0^{(c)} + E_0^{(v)}$, where $E_0^{(c)}$ ($E_0^{(v)}$) represents the electron (hole) ground state energy in the QD, defined with respect to the bulk conduction (valence) band edge in the QD. $E_{binding}$ given above therefore refers to the amount of energy by which $E(e = 0)$ is lowered when 'e' (or the Coulomb interaction among the carriers) is turned on. Note that our definition of $E_{binding}$ differs from the conventional one. The conventional $E_{binding}$ refers to the energy required to separate the trion into a free exciton and a free electron ((in the case of X⁻), which applies in our case only in the limit of infinite QD radius (or the delocalized trion limit) where the exciton-electron system can dissociate into the two far-apart, independent subsystems - an exciton and an electron. In our case, with the two subsystems both being confined in the same QD and constantly interacting with each other, dissociation of the trion can never occur. In the following, we further establish the connection of $E_{binding}$ to optical excitation, from the experimental perspective.

Suppose initially an electron is present in the QD. Then

$$E_{optical}^{(trion)} = E_0^{(c)} + E_0^{(v)} + 2\Delta(r < R) - E_{binding}$$

is the minimal optical excitation energy required to excite across the energy gap an electron-hole pair, which binds with the initial electron into a trion. Here, $2\Delta(r < R)$ is the bulk energy gap inside the QD, and $E_0^{(c)} + E_0^{(v)} + 2\Delta(r < R)$ is the QD energy gap, which is increased from the bulk value due to the quantum confinement. For the optical excitation of an electron-hole pair in the QD, the expression above for $E_{optical}^{(trion)}$ means that the energy required is lowered from the QD gap energy by $E_{binding}$, due to the inter-carrier Coulomb interaction in the QD, when an electron is initially present.

In comparison, in the case where the QD is initially empty, a corresponding discussion would give

$$E_{optical}^{(exciton)} = E_0^{(c)} + E_0^{(v)} + 2\Delta(r < R) - E_{ex},$$

where $E_{optical}^{(exciton)}$ is the minimal energy to excite an electron-hole pair (i.e., exciton) in the QD. Due to the electron-hole Coulomb attraction, $E_{optical}^{(exciton)}$ is reduced from the QD gap energy by the amount $E_{ex}$. $E_{ex}$ is similar to $E_{binding}$ introduced above for a trion, and may thus be defined as the binding energy for an exciton in the QD. Note that with both excitons and trions being analogous few-particle systems, the theory of $E_{ex}$ would be similar to that of $E_{binding}$. For a focused presentation here, such a theory is left to a future, separate work.

The difference between $E_{optical}^{(exciton)}$ and $E_{optical}^{(trion)}$ is given by

$$E_{optical}^{(exciton)} - E_{optical}^{(trion)} = E_{binding} - E_{ex}.$$

The above result establishes the connection between the theory (as represented by the terms on the *r.h.s.*) and the optical measurement (on the *l.f.s.*).

### III. The low-energy trion spectrum and trions in different regimes

**Low-energy trion spectrum**

First, we discuss the low-energy trion spectrum. In particular, we study the total wave function $\Psi_{trion}(r_1, r_2, r_3)$ of trion states with a symmetric envelope function $F_{trion}(r_1, r_2, r_3) = F_{trion}(r_1, r_2, r_3) = F_S(r_1, r_2)$ (with the hole $r_3$-dependence being hidden) and an antisymmetric $pseudospin \otimes spin$ component. These states have been argued in **Sec. I** to be well separated from those having an antisymmetric envelope function $F_{trion}(r_1, r_2, r_3) = -F_{trion}(r_1, r_2, r_3) = F_A(r_1, r_2)$ and a symmetric $pseudospin \otimes spin$ component. We hide the hole part and write $\Psi_{trion}(r_1, r_2, r_3)$ for all the low energy states,

$$\psi_{2e}^{(1)}(\vec{r}_1, \vec{r}_2) = F_S(\vec{r}_1, \vec{r}_2) \psi^{(c)}{}_K(\vec{r}_1) \psi^{(c)}{}_K(\vec{r}_2) \otimes (\text{spin singlet}),$$

$$\psi_{2e}^{(2)}(\vec{r}_1,\vec{r}_2) = F_S(\vec{r}_1,\vec{r}_2)\psi^{(c)}{}_{K'}(\vec{r}_1)\psi^{(c)}{}_{K'}(\vec{r}_2) \otimes (\text{spin singlet}), \tag{17}$$

$$\psi_{2e}^{(3)}(\vec{r}_1,\vec{r}_2) = F_S(\vec{r}_1,\vec{r}_2)\frac{1}{\sqrt{2}}[\psi^{(c)}{}_K(\vec{r}_1)\psi^{(c)}{}_{K'}(\vec{r}_2) + \psi^{(c)}{}_K(\vec{r}_2)\psi^{(c)}{}_{K'}(\vec{r}_1)],$$
$$\otimes (\text{spin singlet})$$

$$\psi_{2e}^{(4)}(\vec{r}_1,\vec{r}_2) = F_S(\vec{r}_1,\vec{r}_2)\frac{1}{\sqrt{2}}[\psi^{(c)}{}_K(\vec{r}_1)\psi^{(c)}{}_{K'}(\vec{r}_2) - \psi_K^{(c)}(\vec{r}_2)\psi^{(c)}{}_{K'}(\vec{r}_1)].$$
$$\otimes (\text{spin triplet})$$

Overall, there are six states in Eqn. (17), counting both spin and pseudospin degeneracy of electrons but discounting that of holes. $\psi_{2e}^{(1)}$ and $\psi_{2e}^{(2)}$ form a doublet. With the two valley pseudospins of electrons being the same in the state, each is a close analogy of the semiconductor trion. $\psi_{2e}^{(3)}$ is nondegenerate, while $\psi_{2e}^{(4)}$ is three-fold degenerate. The pseudospin component of $\psi_{2e}^{(4)}$ is antisymmetric and forms a pseudospin singlet, while those of $\psi_{2e}^{(1)}, \psi_{2e}^{(2)}$, and $\psi_{2e}^{(3)}$ are given, respectively, by

$$\psi^{(c)}{}_K(\vec{r}_1)\psi^{(c)}{}_K(\vec{r}_2),$$

$$\psi^{(c)}{}_{K'}(\vec{r}_1)\psi^{(c)}{}_{K'}(\vec{r}_2),$$

$$\frac{1}{\sqrt{2}}[\psi^{(c)}{}_K(\vec{r}_1)\psi^{(c)}{}_{K'}(\vec{r}_2) + \psi^{(c)}{}_K(\vec{r}_2)\psi^{(c)}{}_{K'}(\vec{r}_1)]$$

which are all symmetric and, altogether, form a set of pseudospin triplet states. However, the pseudospin triplet states are not all degenerate. Due to the symmetric combination of pseudospins, the 3$^{rd}$ one (for $\psi_{2e}^{(3)}$) above has an enhanced probability for $\vec{r}_1 \sim \vec{r}_2$, and thus an increased e-e interaction in comparison to that in the other two. On the other hand, the pseudospin singlet $\psi_{2e}^{(4)}$ has a suppressed probability for $\vec{r}_1 \sim \vec{r}_2$ and thus a reduced e-e interaction. The corresponding energy increase/decrease due to the enhancement/suppression in the probability for $\vec{r}_1 \sim \vec{r}_2$ is an exchange energy, which we discuss and estimate semi-quantitatively below.

**EFA estimation of the trion energy**

We analyze the e-e interaction energy in the 1$^{st}$-order perturbation theory. Two approaches are given here for the estimation. In the first approach, we employ the EFA, yielding

$$E_{e-e}^{(EFA)} = \int |F_S(\vec{r}_1,\vec{r}_2)|^2 \, V_{e-e}(\vec{r}_1-\vec{r}_2) d\vec{r}_1 d\vec{r}_2 \tag{18}$$

for all the low energy states. In the alternative approach, we include the sub-cell details, and replace $F_S(\vec{r}_1,\vec{r}_2)$ by the total wave function $\psi_{2e}^{(n)}(\vec{r}_1,\vec{r}_2)$, n = 1~4. This gives the estimate

$$\begin{aligned}E_{e-e}^{(n)} &= \int |\psi_{2e}^{(n)}(\vec{r}_1,\vec{r}_2)|^2 \, V_{e-e}(\vec{r}_1-\vec{r}_2) d^2\vec{r}_1 d^2\vec{r}_2 \\ &= E_{e-e}\big|_{|\vec{r}_1-\vec{r}_2|\geq O(a)} + E_{e-e}^{(n)}\big|_{|\vec{r}_1-\vec{r}_2|\leq O(a)}, \; n=1\sim 4,\end{aligned} \tag{19}$$

where "$a$" is the lattice constant. The second line above separates the interaction energy integral into the two parts, $E_{e-e}\big|_{|\vec{r}_1-\vec{r}_2|\geq O(a)}$ and $E_{e-e}^{(n)}\big|_{|\vec{r}_1-\vec{r}_2|\leq O(a)}$, depending on the inter-electron distance. Using the fact that in $\psi_{2e}^{(n)}(\vec{r}_1,\vec{r}_2)$, $F_S(\vec{r}_1,\vec{r}_2)$ varies slowly while $\psi^{(c)}_{K(K')}(\vec{r})$ varies rapidly on the length scale ~ $a$, we approximate $E_{e-e}\big|_{|\vec{r}_1-\vec{r}_2|\geq O(a)}$ by first integrating the fast-varying part, yielding, for all the states in Eqn. (17),

$$E_{e-e}\big|_{|\vec{r}_1-\vec{r}_2|\geq O(a)} \approx \int |F_S(\vec{r}_1,\vec{r}_2)|^2 \frac{e^2}{4\pi\varepsilon|\vec{r}_1-\vec{r}_2|} d\vec{r}_1 d\vec{r}_2, $$
$$O(E_{e-e}\big|_{|\vec{r}_1-\vec{r}_2|\geq O(a)}) = \max[e^2/\varepsilon R, e^2/\varepsilon a_B^*]. \tag{18'}$$

$E_{e-e}\big|_{|\vec{r}_1-\vec{r}_2|\geq O(a)}$ obtained here agrees exactly with $E_{e-e}^{(EFA)}$, the EFA estimate given by Eqn. (18). The agreement is consistent with the well-known fact that the EFA applies on the length scale above '$a$'. Within the EFA, the six states in Eqn. (17) are therefore all degenerate. Such a high degeneracy number obviously derives from the presence of valley pseudospin in the trion, apart from spin.

**Beyond the EFA**

However, the states listed in Eqn. (17) do differ in the pseudospin component and, in fact, are split by the short-range e-e exchange interaction energy '$E_{e-e}|_{|\vec{r}_1-\vec{r}_2|\leq a}$' as discussed below. Let $\delta E_{14} = E_{e-e}^{(1)} - E_{e-e}^{(4)} = E_{e-e}^{(2)} - E_{e-e}^{(4)}$, and $\delta E_{34} = E_{e-e}^{(3)} - E_{e-e}^{(4)}$ denote the energy differences. We estimate that

$$\delta E_{14} \approx \frac{1}{2}\delta E_{34}$$

$$\approx \int_{|\vec{r}_1 \sim \vec{r}_2|\leq O(a)} \frac{e^2}{4\pi\varepsilon|\vec{r}_1-\vec{r}_2|}|F_S(\vec{r}_1,\vec{r}_2)|^2 \psi_K(\vec{r}_1)^*\psi_{K'}(\vec{r}_2)^*\psi_K(\vec{r}_2)\psi_{K'}(\vec{r}_1)d\vec{r}_1 d\vec{r}_2, \quad (20)$$

$$O(\delta E_{14}) \approx O(\delta E_{34}) = \frac{e^2}{\varepsilon}\left(\max\left(\frac{a}{R^2},\frac{a}{a_B^{*2}}\right)\right),$$

where 'R' = QD radius. Comparison of the above expression to $\overline{V}_{e-e}^{(K\leftrightarrow K')}(q)$ given in Eqn. (16-b) shows that the energy difference given here derives in origin from the valley-flipping e-e exchange scattering ($H_{e-e}^{(K\leftrightarrow K')}$) discussed in **Sec. II**, which belongs to $H_{ignored}$ and is ignored in the EFA description (i.e., Eqns. (8) and (13)). Therefore, Eqn. (20) can also be regarded as an improvement over the EFA description, through the inclusion of $H_{ignored}$ in the 1st-order perturbation theory. Using Eqn. (20), we can make an order-of-magnitude estimation of the energy splitting. For min(R, $a_B^*$) ~ 100A, it gives $\delta E_{14(34)}$ ~ O (meV). Specifically, since $\{\psi_{2e}^{(1)},\psi_{2e}^{(2)}\}$ are the analogy of the ground state in semiconductors, comparison of their energy to that of $\psi_{2e}^{(4)}$, as represented here by $\delta E_{14}$, provides an interesting assessment of the valley pseudospin involvement in trion energy.

According to Eqns. (18') and (20), it follows that $\delta E_{14(34)} \ll E_{e-e}|_{|\vec{r}_1-\vec{r}_2|\geq O(a)}$ (or $E_{e-e}^{(EFA)}$) for R >> a, meaning that $H_{ignored}$ can be neglected in the limit where R >> a.

### The role of the hole

It would be interesting here to analyze the role of the hole in the trion. This can be done by comparing the trion (with the hole) to the two-electron QD system (without the hole). While the confinement energy ($E_{QD}$), the QD charging energy ($E_{e-e}^{(C)}$) and the exciton binding energy ($E_{ex}$) are all relevant energy scales for the trion, $E_{ex}$ is not for the two-electron system, due to the absence of the hole and the e-h interaction as well from the system.

In addition, it is easy to see that the earlier statement that the $\psi_{2e}(\vec{r}_1,\vec{r}_2)$'s with $F_S(\vec{r}_1,\vec{r}_2)$ are the low-energy states does not necessarily apply to the two-electron system. In the latter case, the ground state for two electrons with opposite pseudospins can be either $F_S(\vec{r}_1,\vec{r}_2)[\psi_K(\vec{r}_1)\psi_{K'}(\vec{r}_2)-\psi_K(\vec{r}_2)\psi_{K'}(\vec{r}_1)]\otimes$(spin triplet) or $F_A(\vec{r}_1,\vec{r}_2)[\psi_K(\vec{r}_1)\psi_{K'}(\vec{r}_2)-\psi_K(\vec{r}_2)\psi_{K'}(\vec{r}_1)]\otimes$(spin singlet), depending on the competition between $E_{QD}$ and $E_{e-e}^{(C)}$ for $|\vec{r}_1-\vec{r}_2|\gtrsim a$, as discussed in the following. First, if $E_{QD}$ dominates, the state with $F_S(\vec{r}_1,\vec{r}_2)$ obviously has the lower energy since it would place the two electrons both in the lowest, one-electron orbital shell of the QD. On the other hand, the state with $F_A(\vec{r}_1,\vec{r}_2)$ tends to place the two electrons in separate shells, thus keeping the electrons apart and reducing the e-e interaction. Therefore, when $E_{e-e}^{(C)}$ dominates, the state with $F_A(\vec{r}_1,\vec{r}_2)$ is favored as the ground state.

### Trions in different regimes

We discuss low energy trions in the various regimes classified using the length scales R and $a_B^*$ or, equivalently, the energy scales $E_{QD}$ (= O($h^2/m^*R^2$)), $E_{e-e}^{(C)}$ (= O($e^2/\varepsilon R$)) and $E_{ex}$ (= O($e^2/\varepsilon a_B^*$)). We define i) the strong confinement regime, where R << $a_B^*$ (or $E_{QD}$ >> $E_{e-e}^{(C)}$ >> $E_{ex}$), ii) the strong interaction regime, where R >> $a_B^*$ (or $E_{QD}$ << $E_{e-e}^{(C)}$ << $E_{ex}$), and iii) the intermediate regime, where R ~ $a_B^*$ (or $E_{QD}$ ~ $E_{e-e}^{(C)}$ ~ $E_{ex}$). Since the present work is mainly concerned with trions in QDs, we focus on the strong confinement regime as well as the intermediate regime, while giving a brief discussion about the strong interaction regime at the end. All regimes are treated in the EFA.

In **the regime of strong confinement**, the trion state can be easily treated with the perturbation theory. For simplicity, we present the symmetric case where $V_e(r) = V_h(r)$ in the following, although the discussion can be easily generalized to the asymmetric case where $V_e(r) \neq V_h(r)$. To the lowest order, we ignore the e-e and the e-h interaction, and consider as examples the lowest trion state with a symmetric $F_S(r_1,r_2,r_3)$ (also the ground state and denoted as $F_0(r_1,r_2,r_3)$) and the lowest state with an antisymmetric $F_A(r_1,r_2,r_3)$ (denoted as $F_1(r_1,r_2,r_3)$). We obtain

$$F_0(r_1,r_2,r_3) \approx \varphi_0^{(c)}(r_1)\varphi_0^{(c)}(r_2)\varphi_0^{(v)}(r_3), \tag{21a}$$

$$F_1(r_1,r_2,r_3) \approx \frac{1}{\sqrt{2}}\left[\varphi_0^{(c)}(r_1)\varphi_1^{(c)}(r_2) - \varphi_0^{(c)}(r_2)\varphi_1^{(c)}(r_1)\right]\varphi_0^{(v)}(r_3) \tag{22b}$$

where $\varphi_0^{(c)}(r)$ and $\varphi_0^{(v)}(r)$ are the one-particle ground states of the confined electron and the confined hole in the QD, respectively, and $\varphi_1^{(c)}(r)$ is the 1$^{st}$ excited electron state. Being bound states, these forgoing functions are all real-valued. Therefore, we have dropped the complex conjugate operation on $\phi_0^{(v)}(r_3)$ in Eqns. (21a) and (21b). Next, we include the e-e and the e-h interaction as the perturbation. The state energies $E_0$ and $E_1$ corresponding to $F_0(r_1,r_2,r_3)$ and $F_1(r_1,r_2,r_3)$ are given, respectively, by the following 1$^{st}$-order perturbation-theoretical expressions

$$\begin{aligned}E_0 &\approx 2E_0^{(c)} + E_0^{(v)} + <\phi_0^{(c)}\phi_0^{(c)}|V_{e-e}|\phi_0^{(c)}\phi_0^{(c)}> \\ &-2<\phi_0^{(c)}\phi_0^{(v)}|V_{e-e}|\phi_0^{(c)}\phi_0^{(v)}>,\end{aligned} \tag{22a}$$

$$\begin{aligned}E_1 &\approx E_0^{(c)} + E_1^{(c)} + E_0^{(v)} + <\varphi_0^{(c)}\varphi_1^{(c)}|V_{e-e}|\varphi_0^{(c)}\varphi_1^{(c)}> \\ &-<\varphi_0^{(c)}\varphi_0^{(v)}|V_{e-e}|\varphi_0^{(c)}\varphi_0^{(v)}> - <\varphi_1^{(c)}\varphi_0^{(v)}|V_{e-e}|\varphi_1^{(c)}\varphi_0^{(v)}> \\ &-<\varphi_0^{(c)}\varphi_1^{(c)}|V_{e-e}|\varphi_1^{(c)}\varphi_0^{(c)}>\end{aligned}. \tag{22b}$$

Here, $E_0^{(c)} = E_0^{(v)}$ (one-particle ground state energy in the QD) and $\varphi_0^{(c)}(r) = \varphi_0^{(v)}(r)$ in the present symmetric case. $E_1^{(c)}$ = one-particle 1$^{st}$ excited state energy in the QD. The matrix element of $V_{e-e}$ above is given by

$$<\varphi_1\varphi_2|V_{e-e}|\varphi_3\varphi_4> \equiv \iint d^2r_1 d^2r_2 [\varphi_1(r_1)\varphi_2(r_2)]^* V_{e-e}(r_1-r_2)\varphi_3(r_1)\varphi_4(r_2).$$

The last term in Eqn. (22b) is the exchange energy between the two electrons, which tends to lower $E_1$.

On the other hand, in **the intermediate regime** where the interaction begins to compete with the quantum confinement, the interaction strongly couples the various product states given by

$$\varphi_{n,m,l}(r_1,r_2,r_3) \equiv \varphi_n^{(c)}(r_1)\varphi_m^{(c)}(r_2)\varphi_l^{(v)}(r_3), \qquad (23)$$

which involves one-particle ground as well as excited states of the QD. Therefore, the solution is generally given by

$$F_{trion}(\vec{r}_1,\vec{r}_2,\vec{r}_3) = \sum_{n,m,l} c_{n,m,l}\varphi_{n,m,l}(\vec{r}_1,\vec{r}_2,\vec{r}_3), \qquad (24)$$

where $\{c_{n,m,l}\text{'s}\}$ are determined by the coupling among the product states. In order to proceed further, we shall resort to the numerical variational calculation as described in the next section.

We briefly remark on the trions **in the strong interaction regime**. In the lowest order treatment, the QD confinement can be ignored, and the trion basically moves freely in the 2D space of graphene. In this approximation, the envelope function $F_{trion}$ is identical to that of the semiconductor trion in a quantum well, when in the latter case we take the thin well limit as well as equal electron and hole masses. This identification yields, in the Schrodinger model, an approximate formula for $E_{binding}$ in the strong interaction regime, which is given by

$$\begin{aligned} E_{binding} &\approx 1.11\, E_{ex}, \\ E_{ex} &= \frac{e^4 \Delta(r<R)}{\hbar^2 v_F^2 \varepsilon^2} \quad (2D\ exciton\ binding\ energy\ in\ gapped\ graphene). \end{aligned} \qquad (25)$$

A similar scaling relation between $E_{binding}$ and $E_{ex}$ (for 3D excitons) is given for semiconductor trions in the thin well limit.[19] Here, it has been adjusted to account for the difference in the definition of $E_{binding}$ between the semiconductor case and ours. Moreover, $E_{binding}$ in Eqn. (25) is expressed in terms of $E_{ex}$ in 2D, which is known to be four times as large as $E_{ex}$ in 3D. [27]

## IV. The variational method

We describe our variational method for solving the trion ground state, in a QD with the following piecewise constant potential profile

$$V_{e(h)}(r) = \begin{cases} -V_{e(h)}^{(0)} & \text{for } r < R, \\ 0 & \text{for } r > R. \end{cases}$$

Since we are primarily interested in the case with R $\gg$ a, it is justified to employ the EFA and calculate the envelope function $F_0(r_1, r_2, r_3)$ for the ground state. Using the wave equation (8) for the envelope function, we write the expectation value of the ground state energy

$$E_0 = <F_0(\vec{r}_1, \vec{r}_2, \vec{r}_3) | H_{trion}^{(model)} | F_0(\vec{r}_1, \vec{r}_2, \vec{r}_3)> \tag{26}$$

subject to the constraint $\int |F_0(\vec{r}_1, \vec{r}_2, \vec{r}_3)|^2 d^2\vec{r}_1 d^2\vec{r}_2 d^2\vec{r}_3 = 1$. $H_{trion}^{(model)}$ in the equation is taken to be $H_{trion}$ given in Eqn. (13), with the mass $m^*(r_i) = \Delta(r_i)/v_F^2$ or $m^*(r_i) = \frac{1}{2}\left[\left(-\hbar^2 \nabla_i^2/v_F^2 + \Delta(r_i)^2/v_F^4\right)^{1/2} + \Delta(r_i)/v_F^2\right]$ defining, respectively, the Schrodinger and Dirac models of trions in gapped graphene.

The variational calculation is performed by expanding $F_0(\vec{r}_1, \vec{r}_2, \vec{r}_3)$ in terms of the following symmetric combination of product states

$$F_0(\vec{r}_1, \vec{r}_2, \vec{r}_3) = \sum_{n,k,l} c_{n,k,l} \phi_{n,k,l}(r_1, r_2, r_3),$$
$$\phi_{n,k,l}(r_1, r_2, r_3) = N_{n,k,l} \left[\chi_n(r_1)\chi_k(r_2) + \chi_n(r_2)\chi_k(r_1)\right]\chi_l(r_3). \tag{27}$$

$N_{n,k,l}$ is the normalization constant. $n = 1 \sim N_1$, $k = 1 \sim N_2$, and $l = 1 \sim N_3$. In a typical calculation, we take $N_1 = N_2 = N_3 \leqq 6$. Each basis $\phi_{n,k,l}(r_1, r_2, r_3)$ is exchange-symmetric with respect to $r_1$ and $r_2$, as required for the ground state.

### **Variational basis functions**

Specifically, we choose four types of functions for $\chi_n(r)$ here, Gaussian, exponential, and two Bessel functions, parameterized by $a_n$, $b_n$, $c_n$, and $d_n$, respectively, as described below. **Appendix A** tabulates $a_n$'s and $b_n$'s (for n = 1 − 6) and $c_n$'s and $d_n$'s (for n = 1 − 5) used in the numerical calculation.

**i)**      **Gaussian functions (for the Schrodinger model)**

The function is parameterized by $a_n$ and given by

$$\chi_n(r) = \begin{cases} \exp[-a_n r^2], & \text{for } r < R \\ \exp[(g_S - 1)a_n R^2]\exp[-g_S a_n r^2], & \text{for } r > R \end{cases} \tag{28}$$

**ii)**      **exponential functions (for the Schrodinger model)**

The function is parameterized by $b_n$ and given by

$$\chi_n(r) = \begin{cases} \exp[-b_n r], & \text{for } r < R \\ \exp[(g_S - 1)b_n R]\exp[-g_S b_n r], & \text{for } r > R \end{cases} \tag{29}$$

**iii)**      **Bessel functions (for the Schrodinger model)**

The function is parameterized by $c_n$ and given by

$$\chi_n(r) = \begin{cases} J_0(c_n r), & \text{for } r < R \\ \alpha_n K_0(c_n' r), & \text{for } r > R \end{cases}. \tag{30}$$

With $K_0$ decaying with increasing r, Eqn. (30) describes a distribution that resembles a bound state. Here, $c_n'$ is a function of $c_n$ determined by

$$\frac{J_0(c_n R)}{K_0(c_n' R)} = \frac{g_S c_n}{c_n'} \frac{J_1(c_n R)}{K_1(c_n' R)},$$

and $\alpha_n$ is a function of $c_n$ and $c_n'$ given by

$$\alpha_n = \frac{J_0(c_n R)}{K_0(c_n' R)}.$$

In Eqns. (28)-(30), $g_S \equiv \frac{m^*(r > R)}{m^*(r < R)}$ being the ratio of effective masses inside and outside the QD, in the Schrodinger model.

iv) **Bessel functions (for the Dirac model)**

The function is parameterized by $d_n$ and given by

$$\chi_n(r) = \begin{cases} J_0(d_n r), & \text{for } r < R \\ \beta_n K_0(d_n' r), & \text{for } r > R \end{cases}. \tag{31}$$

Here, $d_n'$ is a function of $d_n$ determined by

$$\frac{J_0(d_n R)}{K_0(d_n' R)} = g_D(d_n, d_n') \frac{d_n}{d_n'} \frac{J_1(d_n R)}{K_1(d_n' R)},$$

where $g_D(d_n, d_n') \equiv \frac{m^*(r > R; d_n')}{m^*(r < R; d_n)}$ is the ratio of momentum-dependent effective masses inside and outside the QD, which is the generalization of $g_S$ given in the Schrodinger model to the Dirac model. Explicitly, we have

$$m^*(r<R;d_n) = \frac{1}{2}\left[\left(\hbar^2 d_n^2/v_F^2 + \Delta(r<R)^2/v_F^4\right)^{1/2} + \Delta(r<R)/v_F^2\right],$$

$$m^*(r>R;d_n') = \frac{1}{2}\left[\left(-\hbar^2 d_n'^2/v_F^2 + \Delta(r>R)^2/v_F^4\right)^{1/2} + \Delta(r>R)/v_F^2\right].$$

$\beta_n$ is a function of $d_n$ and $d_n'$ given by

$$\beta_n = \frac{J_0(d_n R)}{K_0(d_n' R)}.$$

The spatial widths of $\chi_n(r)$'s in Eqns. (28)-(31) are determined by '$a_n$', '$b_n$', '$c_n$', and '$d_n$', respectively, which are pre-chosen to cover the range $\sim O(a_B^*)$ [or $O(R)$], as listed in **Appendix A**.

**Mass discontinuity and the choice of $\chi_n(r)$**

$\chi_n(r)$'s given above in Eqns. (28) and (29) decay with the constants '$a_n$' or '$b_n$' for $r < R$ and '$g_S a_n$' or '$g_S b_n$' for $r > R$. The reason for the discontinuity in decay constant is discussed below. As mentioned earlier, there is a mass variation, $m^*(r)$, across the QD boundary at $r = R$, giving a non-unity mass ratio $g_S \equiv \frac{m^*(r>R)}{m^*(r<R)}$. The mass variation is taken into account in the trion Hamiltonian given in Eqn. (13), which writes the kinetic energy operator in the form $-\frac{\hbar^2}{2}\nabla_1 \cdot \left(\frac{1}{m^*(r_1)}\nabla_1\right) - \frac{\hbar^2}{2}\nabla_2 \cdot \left(\frac{1}{m^*(r_2)}\nabla_2\right) - \frac{\hbar^2}{2}\nabla_3 \cdot \left(\frac{1}{m^*(r_3)}\nabla_3\right)$. Such an operator requires[24, 26]

$$F_0(r_1,r_2,r_3), \partial_{r_i} F_0(r_1,r_2,r_3)/m^*(r_i) \tag{32}$$

to be continuous across $r_i = R$, for $i = 1$-3. The specification of $\chi_n(r)$ given in Eqns. (28) and (29) enforces the continuity of both $\chi_n(r)$ and $\partial_r \chi_n(r)/m^*(r)$ at $r = R$, as can easily be verified, and therefore ensures the continuity of $F_0(r_1,r_2,r_3)$ and $\partial_{r_i} F_0(r_1, r_2, r_3) / m^*(r_i)$ at $r_i = R$. The Bessel function given in Eqn. (30) is also dictated by the same consideration. Eqn. (31) extends $\chi_n(r)$ defined in Eqn. (30) for the Schrodinger model to the Dirac model, by replacing the mass ratio $g_S$ with $g_D(d_n, d_n')$. In the limit where the gap energy $\Delta$ is predominant (i.e., the Schrodinger regime), $g_D(d_n, d_n')$ reduces to $g_S$ and Eqn. (31) reduces to Eqn. (30).

Eqn. (27) is substituted into Eqn. (26), leading to the variational equation

$$\sum_{n',k',l'} (H_{trion}^{(model)})_{n,k,l;n',k',l'} c_{n',k',l'} = E \sum_{n',k',l'} S_{n,k,l;n',k',l'} c_{n',k',l'},$$
$$(H_{trion}^{(model)})_{n,k,l;n',k',l'} \equiv <\phi_{n,k,l} | H_{trion}^{(model)} | \phi_{n',k',l'}>, \qquad (33)$$
$$S_{n,k,l;n',k',l'} \equiv <\phi_{n,k,l} | \phi_{n',k',l'}>.$$

'S' here is the overlap matrix. Since the basis functions with $\chi_n(r)$'s given by Eqns. (28)-(31) are non-orthogonal, 'S' is different from the identity matrix. **Appendix C** provides the various overlap ($S_{n,k,l;n',k',l'}$'s) and energy integrals ($(H_{trion}^{(model)})_{n,k,l;n',k',l'}$'s) for these basis functions.

The variational calculation proceeds as follows. First, we perform the calculation to study the physics of confined trions in the Schrodinger model, using the three forms of basis functions, namely, Gaussian, exponential, and Bessel functions given in Eqns. (28)-(30). Comparison among the results obtained with the three functions also allows us to assess the numerical reliability of the present variational scheme in general, and that of the calculation using the Bessel function in particular. Next, we employ the Bessel functions given in Eqn. (31) and perform the variational calculation in the Dirac model. In particular, we take the parameter $d_n = c_n$ in Eqn. (31), meaning that we use in the Dirac model the same $J_0$'s that are used in the Schrodinger model, inside the QD. The difference between the Bessel functions in the two models then derives only from the variant in $K_0$ outside the QD, which comes about due to the fact that $K_0$ is connected to $J_0$ in a way depending on the mass ratio, e.g., $g_S$ in the Schrodinger model and $g_D(d_n, d_n')$ in the Dirac model. This ensures a large overlap between the sets of Bessel functions used in the two models. Due to this overlap, similar levels of numerical reliability are expected for the two Bessel function based calculations. Comparison between the two calculations permits us to investigate the "relativistic effect" on the trion state.

## V. Numerical results and discussion

The trion binding energy and wave function in a graphene QD are presented in this section. The QD system is specified by the following parameters: 150 A $\leqq$ R $\leqq$ 450 A, $\varepsilon = 2.4\ \varepsilon_{vacuum}$,[28] $\Delta(r < R) = 28$ meV, $\Delta(r > R) = 84$ meV, $v_F = 1 \times 10^6$ m/sec, electron confinement potential barrier height $V_e^{(0)}$ ($= V_e(r > R) - V_e(r < R)$) = 56 meV, and hole confinement potential barrier height $V_h^{(0)}$ ($= V_h(r > R) - V_h(r < R)$) = 56 meV. These QD structure parameters are used throughout this section unless stated otherwise. With the above parameters, it gives $a_B^* (= \hbar^2 \varepsilon v_F^2 / e^2 \Delta(r < R)) \sim 250$ A for the exciton Bohr radius. With 150 A $< a_B^*$ ($\sim$250 A) $< 450$ A, we expect the trions considered here belong to the intermediate regime.

### The Schrodinger model
Firstly, we discuss the result obtained from the calculation in the Schrodinger model using Gaussian and exponential functions. Using Eqn. (25), we obtain $E_{ex} \sim 23$ meV for the exciton binding energy, and $E_{binding} \sim 25.5$ meV for the trion binding energy in the limit of large R. These values are useful references when we interpret the numerical result.

We calculate the trion binding energy using

$E_{binding} = E(e = 0) - E_0$,

with '$E_0$' here being the ground state energy of trion calculated with the variational method according to Eqn. (33). On the other hand, $E(e = 0)$ above is determined by either of the two methods described in the following.

### $E(e = 0)$ and validity of the variational calculation
$E(e = 0)$ (or the ground state energy of a single confined carrier in the QD) can be calculated numerically with the variational method using the Gaussian and exponential functions specified in Eqns. (28), (29), and **Appendix A**. Apart from the variational method, **Appendix B** provides an analytical solution for the ground state of a confined carrier and, therefore, an alternative method to calculate *E(e = 0)*. In the case where the QD radius is 300 A, the analytical approach yields 16.64 meV as the carrier ground state energy, and the variational method yields 16.67 meV with the Gaussian functions, both with respect to the bottom of QD confinement potential. In the case where the QD radius is 600A, the analytical approach yields 6.50 meV and the variational method yields 6.53 meV with the exponential functions. We further compute $E_{binding}$ variationally using both the Gaussian and exponential functions. In the case where the QD radius is 300 A, it is found that $E_{binding}$ = 37.73 meV with the Gaussian functions and 37.74 meV with the exponential functions. The good agreement between the calculations of both $E_{binding}$ and the single carrier ground state energy indicates a reasonable level of numerical reliability in the variational calculation using the Gaussian and exponential functions.

### The variation of $E_{binding}$ with QD size and confinement potential strength

**Fig. 2** shows how $E_{binding}$ varies with both the electron and the hole confinement potential strength ($V_{e(h)}^{(0)}$), in the case where the QD radius is 300 A. Three barrier heights are considered, namely, $V_{e(h)}^{(0)}$ = 28 meV, 42 meV, and 56 meV. The calculations with Gaussian and exponential functions generally agree well with each other, and they both show that $E_{binding}$ increases with increasing barrier height. In **Fig. 3**, we vary the QD radius 'R' and present the dependence of $E_{binding}$ on R, with R here changing from 150 A to 450 A. It shows that $E_{binding}$ decreases with increasing R due to increasing electron-hole separation, and approaches the free trion limit, 25.5 meV, estimated earlier with Eqn. (25). The calculations with Gaussian, exponential, and Bessel functions agree reasonably well with one another.

### Electron distribution in the trion

**Fig. 4(a)** shows an interesting correlation between the locations of the electrons and the hole, in the case where the QD radius is 300A. In particular, it fixes the hole location at the QD center, and plots the envelope function $F_0(r_1, r_2, r_3 = 0)$, with $F_0$ calculated using both the Gaussian (graph (a)) and exponential (graph (b)) functions. The graphs in (a) and (b) show a strong similarity. For example, both exhibit a discontinuity in the slope across the QD boundary, due to the specific boundary condition in Eqn. (32) for $F_0$. Moreover, with the hole being fixed at the center, $F_0$ reaches the maximum at the high symmetry point ($r_1$ = 0, $r_2$ = 0), and decreases monotonically away from the point. **Fig. 5** further investigates the electron-hole correlation, and shows the electron distribution $F_0(r_1, r_2, r_3 = 300A)$, with the hole now being placed at the edge of the QD. In comparison to Fig. 4, the peak of $F_0$ has shifted away from ($r_1$ = 0, $r_2$ = 0) due to the attractive electron-hole Coulomb interaction, which now pulls the electrons slightly towards the edge where the hole is located.

### Hole distribution in the trion

**Fig. 6** shows the hole distribution $F_0(r_1 = 0, r_2 = 0, r_3)$ in the case where the QD radius is 300A, with the two electrons both being placed at the QD center. $F_0$ is calculated using only the Gaussian functions. With the electrons both sitting at the center, the peak of $F_0$ occurs at ($r_3 = 0$) due to the attractive electron-hole Coulomb interaction. **Fig. 7** shows the hole distribution $F_0(r_1 = 0, r_2 = 300A, r_3)$, with the two electrons now being separately placed at the center and at the edge of the QD. In comparison to Fig. 6, although the present $F_0$ looks similar in shape to that in Fig. 6, it differs, however, in that the amplitude is now overall smaller and less concentrated around the center - while $F_0(r_3=0)/F_0(r_3=300A) = 7.2$ in Fig. 6, $F_0(r_3=0)/F_0(r_3=300A) = 3.8$ in the present case. This shows that spreading the two electrons apart widens the hole distribution as well.

### Numerical reliability in the calculation with Bessel functions

As having been shown in Fig. 3, the variational calculation for trions in the Schrodinger model is also performed using the Bessel functions specified in Eqn. (30) and **Appendix A**, and the result agrees reasonably well with those obtained with the Gaussian or exponential functions. We thus proceed to the variational calculation in the Dirac model, using largely the same Bessel functions (e.g., with $d_n = c_n$) that have already been tested by the calculation.

### The Dirac model

First, we further test the numerical reliability of the Bessel function based variational method, in the case of the Dirac model. For example, we consider the case where the QD radius is 300 A, and calculate the one-particle ground state energy. The variational method yields 16.64 meV with the Bessel functions specified by Eqn. (31) and **Appendix A**, and the analytical approach described in **Appendix B** yields 16.7meV, both with respect to the bottom of QD confinement potential. The forgoing agreement further supports the expectation for a reasonable level of numerical reliability in the variational study of trions, in the case of the Dirac model.

**Fig. 8** presents $E_0^{(c)}$ (ground state energy of the confined electron) as a function of QD radius, derived in both the Schrodinger and Dirac models. **Fig. 9** shows $\varphi_0^{(c)}$ (the corresponding ground state wave function), also derived in both of the models, in the case where the QD radius is 150A. Note that Figs. 8 and 9 also represent $E_0^{(v)}$ and $\varphi_0^{(v)}$, respectively, due to the electron-hole symmetry (i.e., $E_0^{(c)} = E_0^{(v)}$ and $\varphi_0^{(c)} = \varphi_0^{(v)}$). In general, they indicate a lower $E_0^{(c)}$ and a more confined $\varphi_0^{(c)}$ in the Dirac model. However, as shown in the figures, only a slight difference actually exists between the models, indicating that the "relativistic effect" on the confined, one-particle state is small, at least for the class of QDs studied here. A rough explanation is given below in terms of the suppression of the "relativistic effect" by the quantum confinement. First, the "relativistic effect" and the quantum confinement are two phenomena generally prevailing in different regimes, with the former favored in the high energy regime while the latter in the low energy regime. Second, while the effective mass in the Schrodinger model is constant and given by $\Delta/v_F^2$, the mass in the "relativistic" Dirac model is given by $\frac{1}{2}\left[\left(\hbar^2 k^2/v_F^2 + \Delta^2/v_F^4\right)^{1/2} + \Delta/v_F^2\right]$ that increases with the momentum and is always greater than $\Delta/v_F^2$. In the presence of QD confinement, the relatively large "relativistic mass" leads to the lowering of Dirac energy levels in comparison to the Schrodinger ones, and a consequent suppression of the "relativistic effect" due to the energy lowering. In short, the quantum confinement limits the "relativistic effect" and results in the demonstrated proximity between the two models in Figs. 8 and 9. In the following, we compare the two models in the case of trions, where the inter-particle Coulomb interaction comes into play.

**Fig. 10** presents the trion binding energy $E_{binding}$ as a function of QD radius, calculated in both the Schrodinger and Dirac models. Generally, $E_{binding}$ is slightly larger in the Dirac model. **Fig. 11** shows $F_0(r_1, r_2, r_3 = 0)$ calculated in the two models, in the case where the QD radius is 150A. It is found that $F_0$ is more concentrated near the QD center in the Dirac model. In both figures, the contrast between the two models can be attributed to the enhanced Dirac carrier confinement in the QD shown in Fig. 9. Due to the enhanced confinement, it effectively increases, in the Dirac model, the electron-hole attraction and, hence, the electron-hole correlation. In Fig. 11 where the hole location is fixed at the center ($r_3 = 0$), this results in the enhanced $F_0$ near $r_1 = r_2 = 0$ in Fig. 11(b) (for the Dirac model), when compared to Fig. 11(a) (for the Schrodinger model). On the other hand, due to the enhanced carrier confinement, both the electron-electron repulsion and the electron-hole attraction are effectively raised. However, they offset each other largely in their effects on the trion energy ($E_0$). Overall, the offset leads, along with the quantum suppression of the "relativistic effect" on the one-particle energy (or E(e=0)), to the limited, net "relativistic effect" on $E_{binding}$ shown in Fig. 10.

**A note on the Dirac model**

The Dirac model employed in the variational calculation is, rigorously speaking, an approximate theory of Dirac trions. Basically, as expressed in Eqns. (8) and (13), it employs the exact Dirac energy dispersion, but treats both electron and hole states approximately as one-band states. Such an approximation lowers the numerical complexity level of the calculation considerably, down to that of the Schrodinger model. On the other hand, the reduced, one-band description could, in principle, be improved by utilizing the full two-band Dirac theory [3] of electron and hole states. **Appendix D** discusses such a two-band model of trions, and compares the present reduced description of trions to the two-band model. Based on the comparison, it also gives an assessment of the reduced model for the class of quantum dots considered here.

## VI. Summary

We have investigated, within the envelope function approximation, the low-energy states of a trion in a graphene quantum dot. The presence of valley pseudospin adds convolution to the interplay between exchange symmetry and the e-e interaction in the trion. The involvement of valley pseudospin leads to new states of trions different from those in semiconductors, and it is found that the low energy trion states are nearly degenerate, and form either spin singlet or triplet, as opposed to the nondegenerate, spin singlet-only ground state in a semiconductor. We have performed the study of trions analytically as well as numerically, with the numerical work being carried out within the variational method using simple products of Gaussian or exponential functions as basis. Calculations with the two types of functions are found to agree well with each other. The trion binding energy is found to decrease with increasing QD size and increase with increasing confinement potential strength. Electron and hole distributions have been investigated, and interesting correlation between electron and hole locations has been demonstrated. "Relativistic effects" have also been investigated with the variational method using the Bessel functions. In the study of electron-hole correlation in trions, a sizable enhancement due to the "relativistic effect" is demonstrated. On the other hand, in the study of trion binding energy, a close proximity between the Schrodinger and Dirac models is found, which is attributed to the strong suppression of the "relativistic effect" by quantum confinement, as well as the competition between the electron-electron repulsion and the electron-hole attraction, in the case of trion binding energy.

Trion states are relatively simple and, hence, can reflect plainly the effects of both quantum confinement and the interaction between charge carriers in the QD. In particular, a large $E_{binding}$ ~ a few tens of meV has been demonstrated in our numerical work, which implies that these effects are reasonably easy to observe in experiments on trions. Therefore, apart from a revelation of the interesting role of valley pseudospin in the formation of low-energy trion states, we expect that this theoretical study will aid the class of experiments which probe these bound states as a way to characterize graphene nanostructures or investigate the few-electron interaction physics in these structures.

**Acknowledgment** – We would like to thank the support of ROC National Science Council through the Contract No. NSC102-2112-M-007-013.

# Appendix A: Parameters of basis functions

We list $a_n$'s, $b_n$'s (n = 1-6), $c_n$'s and $d_n$'s (n = 1-5) of the basis functions used in the variational calculation.

*Gaussian*

| R | $(a_1)^{-1/2}$ | $(a_2)^{-1/2}$ | $(a_3)^{-1/2}$ | $(a_4)^{-1/2}$ | $(a_5)^{-1/2}$ | $(a_6)^{-1/2}$ |
|---|---|---|---|---|---|---|
| 150 Å | 65 Å | 130 Å | 195 Å | 260 Å | 325 Å | 390 Å |
| 200 Å | 65 Å | 130 Å | 195 Å | 260 Å | 325 Å | 390 Å |
| 250 Å | 65 Å | 130 Å | 195 Å | 260 Å | 325 Å | 390 Å |
| 300 Å | 65 Å | 130 Å | 195 Å | 260 Å | 325 Å | 390 Å |
| 350 Å | 80 Å | 160 Å | 240 Å | 320 Å | 360 Å | 480 Å |
| 400 Å | 80 Å | 160 Å | 240 Å | 320 Å | 360 Å | 480 Å |
| 450 Å | 80 Å | 160 Å | 240 Å | 320 Å | 360 Å | 480 Å |

*Exponential*

| R | $(b_1)^{-1}$ | $(b_2)^{-1}$ | $(b_3)^{-1}$ | $(b_4)^{-1}$ | $(b_5)^{-1}$ | $(b_6)^{-1}$ |
|---|---|---|---|---|---|---|
| 150 Å | 12.429 Å | 33.281 Å | 64.2 Å | 105.185 Å | 156.237 Å | 217.356 Å |
| 200 Å | 10.272 Å | 33.281 Å | 69.439 Å | 118.744 Å | 181.198 Å | 256.8 Å |
| 250 Å | 20.133 Å | 49.716 Å | 92.448 Å | 148.328 Å | 217.356 Å | 299.532 Å |
| 300 Å | 20.133 Å | 49.716 Å | 92.448 Å | 148.328 Å | 217.356 Å | 299.532 Å |
| 350 Å | 20.133 Å | 49.716 Å | 92.448 Å | 148.328 Å | 217.356 Å | 299.532 Å |
| 400 Å | 20.133 Å | 49.716 Å | 92.448 Å | 148.328 Å | 217.356 Å | 299.532 Å |
| 450 Å | 20.133 Å | 49.716 Å | 92.448 Å | 148.328 Å | 217.356 Å | 299.532 Å |

*Bessel*

| R | $(c_1)^{-1},(d_1)^{-1}$ | $(c_2)^{-1},(d_2)^{-1}$ | $(c_3)^{-1},(d_3)^{-1}$ | $(c_4)^{-1},(d_4)^{-1}$ | $(c_5)^{-1},(d_5)^{-1}$ |
|---|---|---|---|---|---|
| 150 Å | 208.02 Å | 167.85 Å | 143.17 Å | 115.89 Å | 97.35 Å |
| 200 Å | 486.76 Å | 290.89 Å | 208.02 Å | 167.85 Å | 143.17 Å |
| 250 Å | 685.86 Å | 486.76 Å | 290.89 Å | 208.02 Å | 167.85 Å |
| 300 Å | 685.86 Å | 486.76 Å | 290.89 Å | 208.02 Å | 167.85 Å |
| 350 Å | 685.86 Å | 486.76 Å | 347.69 Å | 290.89 Å | 208.02 Å |
| 400 Å | 685.86 Å | 486.76 Å | 347.69 Å | 290.89 Å | 221.26 Å |
| 450 Å | 685.86 Å | 486.76 Å | 347.69 Å | 290.89 Å | 243.38 Å |

**Appendix B: One-carrier ground state**

We present the exact one-carrier ground state in the two-dimensional QD when a gap or mass discontinuity is present. The problem is treated in both the Schrodinger and the Dirac models of graphene.

**1) The Schrodinger model**

In a type-I graphene QD, the effective Schrodinger equation for a single electron or hole is given by[10]

$$-\frac{\hbar^2}{2}\nabla \cdot \frac{1}{m^*(r)}\nabla\Psi + V_{e(h)}(r)\Psi = E\Psi \qquad \text{(B-1)}$$

where the subscript "e" stands for electron and "h" for hole, and we have taken the energy zero to be at the corresponding band edge outside the QD. In the equation,

$$\frac{m^*(r>R)}{m^*(r<R)} = g_S,$$

$$V_{e(h)}(r) = \begin{cases} -V_{e(h)}^{(0)} & \text{for } r < R, \\ 0 & \text{for } r > R. \end{cases}$$

We introduce the following dimensionless variables,

$$r' \equiv r/R, \ \omega \equiv E/\varepsilon_0, \ \mathrm{m}'(r') \equiv m^*(r)/m^*(r<R), \ V'(r') \equiv V_{e(h)}(r)/\varepsilon_0,$$

where $\varepsilon_0 \equiv \dfrac{\hbar^2}{2m^*(r<R)R^2}$. Then (B-1) reduces to

$$-\nabla' \cdot \frac{1}{m'(r')}\nabla'\Psi + V'(r')\Psi = \omega\Psi \qquad \text{(B-2)}$$

where

$$m'(r') = \begin{cases} 1 & \text{for } r' < 1 \\ g_S & \text{for } r' > 1 \end{cases},$$

$$V'(r') = \begin{cases} -V_{e(h)}^{(0)}/\varepsilon_0 & \text{for } r' < 1 \\ 0 & \text{for } r' > 1 \end{cases}.$$

(B-2) can be solved for the ground state in terms of Bessel functions, yielding

$$\Psi(r') = \begin{cases} J_0(\sqrt{\omega + V_{e(h)}^{(0)}/\varepsilon_0}\, r'), & 0 < r' < 1 \\ \alpha K_0(\sqrt{-g_S \omega}\, r'), & r' > 1 \end{cases} \quad \text{(B-3)}$$

Here, $\alpha$ and $\omega$ are to be determined by the following continuity conditions at r' = 1:

$$\Psi(1^-) = \Psi(1^+),$$
$$\frac{d\Psi}{dr'}\bigg|_{r'=1^-} = \frac{1}{g_S}\frac{d\Psi}{dr'}\bigg|_{r'=1^+} \quad \text{(B-4)}$$

Using the following identities for Bessel functions

$$\frac{dJ_0(z)}{dz} = -J_1(z) \qquad \frac{dK_0(z)}{dz} = -K_1(z),$$

(B-4) leads to

$$\frac{J_0(\sqrt{\omega + V_{e(h)}^{(0)}/\varepsilon_0})}{K_0(\sqrt{-g_S\omega})} = \frac{g_S\sqrt{\omega + V_{e(h)}^{(0)}/\varepsilon_0}}{\sqrt{-g_S\omega}} \frac{J_1(\sqrt{\omega + V_{e(h)}^{(0)}/\varepsilon_0})}{K_1(\sqrt{-g_S\omega})}, \quad \text{(B-5a)}$$

$$\alpha = \frac{J_0(\sqrt{\omega + V_{e(h)}^{(0)}/\varepsilon_0})}{K_0(\sqrt{-g_S\omega})}. \quad \text{(B-5b)}$$

(B-5a) is a transcendental equation satisfied by the dimensionless energy eigenvalue ω, and its root can be found numerically to yield ω. (B-5b) provides $\alpha$ once $\omega$ is obtained. Therefore, this solves the one-particle ground state in the QD.

**2) The Dirac model**

The Dirac theory of graphene is a two-band model consisting of two 1st-order differential equations.[3] We merge the two equations, and in the case of a type-I graphene QD, this yields the equation for a single electron or hole[10]

$$-\frac{\hbar^2}{2}\nabla\cdot\frac{1}{m^*(r;E)}\nabla\Psi + V_D(r)\Psi = E\Psi \tag{B-6}$$

where

$$m^*(r;E) \equiv \frac{E + 2\Delta(r) - V_D(r)}{2v_F^2},$$

$$V_D(r) = \begin{cases} -V_{e(h)}^{(0)} & \text{for } r < R, \\ 0 & \text{for } r > R. \end{cases}$$

("$e$" for an electron and "$h$" for a hole)

The energy zero here is taken to be at the corresponding band edge outside the QD. Eqn. (B-6) appears to be of the same form as Eqn. (B-1). However, the function $m^*(r;E)$ introduced here, which plays the role of "effective mass" in the equation, is energy-dependent, as it should be in a Dirac type theory.

We define the energy-dependent mass ratio $g_D(E)$

$$g_D(E) = \frac{m^*(r>R;E)}{m^*(r<R;E)},$$

and also introduce the following dimensionless variables,

$$r' \equiv r/R,\ \omega_D(E) \equiv E/\varepsilon_{0,D}(E),\ m'(r';E) \equiv m^*(r;E)/m^*(r<R;E),$$
$$V_D'(r';E) \equiv V_D(r)/\varepsilon_{0,D}(E)$$

where $\varepsilon_{0,D}(E) \equiv \dfrac{\hbar^2}{2m^*(r<R;E)R^2}$. Then (B-6) reduces to

$$-\nabla'\cdot\frac{1}{m'(r';E)}\nabla'\Psi + V_D'(r';E)\Psi = \omega_D(E)\Psi \tag{B-7}$$

where

$$m'(r';E) = \begin{cases} 1 & \text{for } r' < 1 \\ g_D(E) & \text{for } r' > 1 \end{cases},$$

$$V_D'(r';E) = \begin{cases} -V_{e(h)}^{(0)} / \varepsilon_{0,D}(E) & \text{for } r' < 1 \\ 0 & \text{for } r' > 1 \end{cases}.$$

(B-7) can be solved for the ground state in terms of Bessel functions, yielding

$$\Psi(r') = \begin{cases} J_0(\sqrt{\omega_D(E) + V_{e(h)}^{(0)}/\varepsilon_{0,D}(E)}\, r'), & 0 < r' < 1 \\ \beta K_0(\sqrt{-g_D(E)\omega_D(E)}\, r'), & r' > 1 \end{cases} \quad \text{(B-8)}$$

The unknowns $\beta$ and E in the equation are to be determined by the following continuity conditions at r' = 1:

$$\Psi(1^-) = \Psi(1^+),$$

$$\frac{d\Psi}{dr'}\bigg|_{r'=1^-} = \frac{1}{g_D}\frac{d\Psi}{dr'}\bigg|_{r'=1^+}$$

or

$$\frac{J_0(\sqrt{\omega_D(E) + V_{e(h)}^{(0)}/\varepsilon_{0,D}(E)})}{K_0(\sqrt{-g_D(E)\omega_D(E)})}$$
$$= \frac{g_D(E)\sqrt{\omega_D(E) + V_{e(h)}^{(0)}/\varepsilon_{0,D}(E)}}{\sqrt{-g_D(E)\omega_D(E)}} \frac{J_1(\sqrt{\omega_D(E) + V_{e(h)}^{(0)}/\varepsilon_{0,D}(E)})}{K_1(\sqrt{-g_D(E)\omega_D(E)})}, \quad \text{(B-9a)}$$

$$\beta = \frac{J_0(\sqrt{\omega_D(E) + V_{e(h)}^{(0)}/\varepsilon_{0,D}(E)})}{K_0(\sqrt{-g_D(E)\omega_D(E)})}. \quad \text{(B-9b)}$$

(B-9a) is a transcendental equation satisfied by E, and its root can be found numerically. (B-9b) provides $\beta$ once E is obtained. Therefore, this solves the one-particle ground state in the QD.

The solution derived here allows us to determine, in either the Schrodinger or the Dirac model, the ground state energy of the noninteracting system of two electrons and one hole, giving *E(e = 0)* as a reference energy for the calculation of the trion binding energy $E_{binding}$.

**Appendix C: Useful integrals for the variational calculation**

In the following, we provide the integrals involved in both the Hamiltonian and the overlap matrix elements in Eqn. (33).

**i)    Gaussian and exponential functions**

We introduce the notations

$$<r|n> = \begin{cases} <r|n>_< = e^{-nr^2}, & r < R \\ <r|n>_> = e^{n(g_S-1)R^2} e^{-g_S nr^2}, & r > R \end{cases}$$

$$<r_1 r_2 r_3 | nkl > = \begin{cases} <r_1 r_2 r_3 | nkl >_< = e^{-(nr_1^2 + kr_2^2 + lr_3^2)}, & r < R \\ <r_1 r_2 r_3 | nkl >_> = e^{(n+k+l)(g_S-1)R^2} e^{-g_S(nr_1^2 + kr_2^2 + lr_3^2)}, & r > R \end{cases}$$

for the Gaussian function, and

$$<r|n> = \begin{cases} <r|n>_< = e^{-nr}, & r < R \\ <r|n>_> = e^{n(g_S-1)R} e^{-g_S nr}, & r > R \end{cases}$$

$$<r_1 r_2 r_3 | nkl > = \begin{cases} <r_1 r_2 r_3 | nkl >_< = e^{-(nr_1 + kr_2 + lr_3)}, & r < R \\ <r_1 r_2 r_3 | nkl >_> = e^{(g_S-1)(n+k+l)R} e^{-g_S(nr_1 + kr_2 + lr_3)}, & r > R \end{cases}$$

for the exponential function. '$g_S$' in the above is the mass ratio (= $m^*(r > R) / m^*(r < R)$).

**The overlap integral** between two basis functions is given by

$$\langle n|n'\rangle = \langle n|n'\rangle_< + \langle n|n'\rangle_>$$

$$= \begin{cases} 2\pi \left[ \dfrac{1 - e^{-(n+n')R^2}}{2(n+n')} + \dfrac{e^{-(n+n')R^2}}{2(n+n')g_S} \right] & \text{(Gaussian basis)}, \\ 2\pi \left[ \dfrac{1 - [1+(n+n')R]e^{-(n+n')R}}{(n+n')^2} + \dfrac{[1+(n+n')g_S R]e^{-(n+n')R}}{(n+n')^2 g_S^2} \right] & \text{(exponential basis)}. \end{cases} \quad (C-1)$$

**The kinetic energy** in the Hamiltonian matrix element involves the integral

$$\langle n|-\nabla\cdot\left(\frac{1}{m^*}\nabla\right)|n'\rangle = \frac{1}{m^*(r<R)}\left[\langle n|-\nabla^2|n'\rangle_< + \frac{1}{g}\langle n|-\nabla^2|n'\rangle_>\right]$$

$$= \begin{cases} \dfrac{2\pi}{m^*(r<R)}\left\{\dfrac{2n'(1-e^{-(n+n')R^2})}{(n+n')} - \dfrac{2n'^2\{1-[1+(n+n')R^2]e^{-(n+n')R^2}\}}{(n+n')^2} \right. \\ \left. + \dfrac{e^{-(n+n')R^2}}{g_S}\left[\dfrac{2n'}{(n+n')} - \dfrac{2n'^2[1+(n+n')g_S R^2]}{(n+n')^2}\right]\right\} \\ \text{(Gaussian basis),} \\[1em] \dfrac{2\pi}{m^*(r<R)}\left\{\dfrac{n'(1-e^{-(n+n')R})}{(n+n')} - \dfrac{n'^2\{1-[1+(n+n')R]e^{-(n+n')R}\}}{(n+n')^2} \right. \\ \left. + \dfrac{e^{-(n+n')R}}{g_S}\left[\dfrac{n'}{(n+n')} - \dfrac{n'^2[1+(n+n')g_S R]}{(n+n')^2}\right]\right\} \\ \text{(exponential basis).} \end{cases} \quad (C\text{-}2)$$

**The Coulomb interaction energy** in the Hamiltonian matrix element involves the integral

$$\langle nkl|\frac{1}{r_{12}}|n'k'l'\rangle$$

$$= \begin{cases} (2\pi)^2\left[\dfrac{1-e^{-(l+l')R^2}}{2(l+l')} + \dfrac{e^{-(l+l')R^2}}{2(l+l')g_S}\right]\times \\ \displaystyle\int_0^\infty dr_1\int_0^\infty dr_2\int_0^{2\pi}d\theta_{12}\dfrac{r_1 r_2 <n|r_1><k|r_2><r_1|n'><r_2|k'>}{\sqrt{r_1^2+r_2^2-2r_1 r_2\cos\theta_{12}}} \text{ (Gaussian basis),} \\[1em] (2\pi)^2\left[\dfrac{1-[1+(l+l')R]e^{-(l+l')R}}{(l+l')^2} + \dfrac{[1+(l+l')g_S R]e^{-(l+l')R}}{(l+l')^2 g_S^2}\right]\times \\ \displaystyle\int_0^\infty dr_1\int_0^\infty dr_2\int_0^{2\pi}d\theta_{12}\dfrac{r_1 r_2 <n|r_1><k|r_2><r_1|n'><r_2|k'>}{\sqrt{r_1^2+r_2^2-2r_1 r_2\cos\theta_{12}}} \text{ (exponential basis).} \end{cases} \quad (C\text{-}3)$$

The above three-dimensional integrals are numerically evaluated in our work.

**The QD confinement potential energy** in the Hamiltonian matrix element involves the integral

$$\langle n|V_{e(h)}(r)|n'\rangle = \begin{cases} -\dfrac{\pi V_{e(h)}^{(0)}}{(n+n')}\left(1-e^{-(n+n')R^2}\right), \text{ (Gaussian basis)} \\ -\dfrac{2\pi V_{e(h)}^{(0)}}{(n+n')^2}\left\{1-[1+(n+n')R]e^{-(n+n')R}\right\}, \text{ (exponential basis)} \end{cases} \quad (C\text{-}4)$$

where we have taken $V_{e(h)}(r) = 0$ outside the QD, and $V_{e(h)}(r) = -V_{e(h)}^{(0)}$ inside the QD.

### ii) Bessel functions

We introduce the notations

$$<r|n> = \begin{cases} <r|n>_< = J_0(c_n r), \ r<R \\ <r|n>_> = \alpha_n K_0(c_n' r), \ r>R \end{cases} \text{ (for the Schrodinger model)}$$

$$<r|n> = \begin{cases} <r|n>_< = J_0(d_n r), \ r<R \\ <r|n>_> = \beta_n K_0(d_n' r), \ r>R \end{cases} \text{ (for the Dirac model)}$$

$$<r_1 r_2 r_3|nkl> = \begin{cases} <r_1 r_2 r_3|nkl>_< = <r_1|n>_< <r_2|k>_< <r_3|l>_<, \ r<R \\ <r_1 r_2 r_3|nkl>_> = <r_1|n>_> <r_2|k>_> <r_3|l>_>, \ r>R \end{cases}$$

**The overlap integral** between two basis functions is given by

$$\langle m|n\rangle = \langle m|n\rangle_< + \langle m|n\rangle_> = 2\pi\left[\int_0^R rdr <m|r>_< <r|n>_< + \int_R^\infty rdr <m|r>_> <r|n>_>\right]$$

$$\int_0^R rdr <m|r>_< <r|n>_< = \begin{cases} \dfrac{R}{c_n^2-c_m^2}[c_n J_0(c_m R)J_1(c_n R) - c_m J_0(c_n R)J_1(c_m R)], \\ \text{(Schrodinger model)} \\ \dfrac{R}{d_n^2-d_m^2}[d_n J_0(d_m R)J_1(d_n R) - d_m J_0(d_n R)J_1(d_m R)], \\ \text{(Dirac model)} \end{cases} \quad (C\text{-}5)$$

$$\int_R^\infty rdr <m|r>_> <r|n>_>$$

$$= \begin{cases} \dfrac{\alpha_m \alpha_n R}{c_n'^2 - c_m'^2}[c_n' K_0(c_m' R)K_1(c_n' R) - c_m' K_0(c_n' R)K_1(c_m' R)], \\ \text{(Schrodinger model)} \\ \dfrac{\beta_m \beta_n R}{d_n'^2 - d_m'^2}[d_n' K_0(d_m' R)K_1(d_n' R) - d_m' K_0(d_n' R)K_1(d_m' R)]. \\ \text{(Dirac model)} \end{cases} \quad \text{(C-6)}$$

**The sum of the kinetic energy and the QD confinement potential energy** gives the following Hamiltonian matrix element

$$\left\langle m \left| -\frac{\hbar^2}{2} \nabla \cdot \left(\frac{1}{m^*}\nabla\right) + V_{e(h)} \right| n \right\rangle$$

$$= \begin{cases} \left(\dfrac{\hbar^2 c_n^2}{2m^*(r<R)} - V_{e(h)}^{(0)}\right)\langle m|n\rangle_< - \left(\dfrac{\hbar^2 c_n'^2}{2m^*(r>R)}\right)\langle m|n\rangle_>, \\ \text{(Schrodinger model)} \\ \left[\left(\hbar^2 v_F^2 d_n^2 + \Delta^2\right)^{1/2} - V_{e(h)}^{(0)}\right]\langle m|n\rangle_< + \left(-\hbar^2 v_F^2 d_n'^2 + \Delta^2\right)^{1/2}\langle m|n\rangle_> \\ -\Delta(r<R)\langle m|n\rangle_< - \Delta(r>R)\langle m|n\rangle_>. \\ \text{(Dirac model)} \end{cases} \quad \text{(C-7)}$$

Eqn. (C-7) involves the integrals $\langle m|n\rangle_<$ and $\langle m|n\rangle_>$, both of which have already been evaluated in Eqns. (C-5) and (C-6).

**The Coulomb interaction energy** in the Hamiltonian matrix element involves integral

$$\left\langle nkl \left| \frac{1}{r_{12}} \right| n'k'l' \right\rangle$$
$$= 2\pi \langle l|l'\rangle \int_0^\infty dr_1 \int_0^\infty dr_2 \int_0^{2\pi} d\theta_{12} \frac{r_1 r_2 <n|r_1><k|r_2><r_1|n'><r_2|k'>}{\sqrt{r_1^2 + r_2^2 - 2r_1 r_2 \cos\theta_{12}}} \quad \text{(C-8)}$$

(for both the Schrodinger and Dirac models)

The above three-dimensional integral is numerically evaluated in our work.

# Appendix D: The reduced trion model vs. the full two-band trion model in the Dirac regime

The reduced Dirac model given in Eqns. (8) and (13) replaces the quadratic energy dispersion in the Schrodinger model by the Dirac dispersion, but treats the one-carrier states approximately. For example, for the states around the Dirac point, it expresses them as $e^{ik\cdot r}\psi^{(c)}_{K(K')}(r)$ and $e^{ik\cdot r}\psi^{(v)}_{K(K')}(r)$ with $\psi^{(c)}_{K(K')}(r)$ and $\psi^{(v)}_{K(K')}(r)$ being the wave functions at the Dirac point, as indicated in Eqn. (3). In the following, we consider the full, two-band Dirac model that includes the more exact wave functions, for example, the ones obtained by directly solving the 2x2 Dirac equation. Such a two-band description would allow for a suitable treatment of the mixing between conduction and valence band states in trions.

First, we introduce the one-carrier state in the two-band Dirac theory. The wave function of a conduction band electron around, for example, the K point is given by $\frac{1}{\sqrt{\Omega}}e^{ik\cdot r}\psi^{(c)}_{k+K}(r)$ (k = wave vector), with

$$\psi^{(c)}_{k+K}(r) = e^{iK\cdot r}U^{(c)}_{k+K}(\vec{r}),$$

$$U^{(c)}_{k+K}(\vec{r}) = \begin{pmatrix} c_k^{(A)} u_K^{(A)}(r) \\ c_k^{(B)} u_K^{(B)}(r) \end{pmatrix}. \tag{D-1}$$

$u_K^{(A)}(r)$ and $u_K^{(B)}(r)$ are cell-periodic Bloch functions. The superscripts 'A' and 'B' indicate the two atomic sites in a graphene unit cell. The coefficients $c_k^{(A)}$ and $c_k^{(B)}$ represent, respectively, the amplitudes on 'A' and 'B' sites. They are determined by the 2 x 2 Dirac equation and given by

$$\begin{pmatrix} c_k^{(A)} \\ c_k^{(B)} \end{pmatrix} \propto \begin{pmatrix} E + 2\Delta \\ \hbar v_F k_+ \end{pmatrix}, \tag{D-2}$$

where $k_+ = k_x + ik_y$ and 'E' = the corresponding electron energy with respect to the conduction band edge. The amplitudes satisfy the normalization condition

$$|c_k^{(A)}|^2 + |c_k^{(B)}|^2 = 1.  \tag{D-3}$$

We take 'A' site to have the higher atomic energy level 'Δ', and "B" site the lower energy level '- Δ', where Δ = energy gap parameter of graphene. It is known that a conduction band electron mainly occupies 'A' sites while a valence band electron mainly occupies 'B' sites. Therefore, for a conduction band state, $(c_k^{(A)}, c_k^{(B)}) \approx (1,0)$, and the amplitude $|c_k^{(B)}|^2$ increases with increasing electron momentum, reflecting the "relativistic effect'. In the reduced trion model, we ignore the state mixing and take

$$(c_k^{(A)}, c_k^{(B)}) = (1,0).  \tag{D-4}$$

Eqns. (D-1) – (D-4) can be generalized to valence electron states and states round the K' valley.

Next, we compare the reduced model to the full two-band model of trions, as follows.

a) First, we discuss the kinetic aspect. We consider the central trion envelope function equation in the reduced model, specified by Eqns. (8) and (13), in the noninteracting limit. In this case, the equation reduces to the wave equation for three independent particles, with each particle governed by the same one-band Dirac equation in **Appendix B**. Moreover, since the one-band Dirac equation there is rigorously derived from the 2 x 2 Dirac equation (by combining into one the two differential equations in the 2 x 2 problem), the reduced model thus exactly conforms to the 2 x 2 Dirac theory, in the noninteracting case.

b) Second, we discuss the interaction aspect. We turn on the inter-carrier interaction in the trion, and estimate the difference between the reduced and the full Dirac models, in the interaction matrix element.

We consider a typical matrix element for the interaction between, for example, two electrons - one K valley electron (with wave vectors $k_1$ and $k_1'$) and the other K' valley electron (with wave vectors $k_2$ and $k_2'$). This is a case relevant to the trion (consisting of one K electron, one K' electron, and one K' hole) considered in **Sec. II**.

In the two-band Dirac model, we would have

$$<k_1, k_2 | V_{e-e} | k_1', k_2'>$$
$$= \frac{1}{\Omega^2} \int d^2 r_1 d^2 r_2 \, e^{-i(\vec{k}_1 - \vec{k}_1') \cdot (\vec{r}_1 - \vec{r}_2)} U_{k_1+K}^{(c)*}(\vec{r}_1) U_{k_2+K'}^{(c)*}(\vec{r}_2) V_{e-e}(\vec{r}_1 - \vec{r}_2)$$
$$U_{k_1'+K}^{(c)}(\vec{r}_1) U_{k_2'+K'}^{(c)}(\vec{r}_2)$$
$$\approx \overline{V}_{e-e}(\vec{k}_1 - \vec{k}_1') \left( c_{k_1}^{(A)*} c_{k_1'}^{(A)} + c_{k_1}^{(B)*} c_{k_1'}^{(B)} \right) \left( d_{k_2}^{(A)*} d_{k_2'}^{(A)} + d_{k_2}^{(B)*} d_{k_2'}^{(B)} \right)$$
(for the two-band model)  (D-5)

where the constants $d_{k_2}^{(A)}$, $d_{k_2'}^{(A)}$, $d_{k_2}^{(B)}$ and $d_{k_2'}^{(B)}$ represent the amplitudes of the K' electron on 'A' sites and 'B' sites, respectively.

In the reduced Dirac model, with the approximation in Eqn. (D-4),

$$<k_1, k_2 | V_{e-e} | k_1', k_2'> \approx \overline{V}_{e-e}(\vec{k}_1 - \vec{k}_1')$$
(for the reduced model)  (D-6)

If we compare Eqns. (D-5) and (D-6), we see that the difference is due to the presence of a non-unity factor in Eqn. (D-5), given by

$$\mu \equiv \left( c_{k_1}^{(A)*} c_{k_1'}^{(A)} + c_{k_1}^{(B)*} c_{k_1'}^{(B)} \right) \left( d_{k_2}^{(A)*} d_{k_2'}^{(A)} + d_{k_2}^{(B)*} d_{k_2'}^{(B)} \right) \quad \text{(D-7)}$$

$\mu = 1$ in the reduced model, and deviates from unity in the two-band model, because the amplitudes on site 'B' are generally nonvanishing for states away from the Dirac point (i.e., state mixing). Thus, as shown above, the state mixing results in a modification of the interaction matrix element.

Now, we make an estimate of the deviation of $\mu$ from unity, in a typical quantum dot, in the two-band model. Specifically, we take the radius to be 300A. We consider one of the inner products, for example, $\rho = c_{k_1}^{(A)*} c_{k_1'}^{(A)} + c_{k_1}^{(B)*} c_{k_1'}^{(B)}$ in Eqn. (D-7). The task then is to evaluate the average of ρ, with $(c_{k_1}^{(A)}, c_{k_1}^{(B)})$ and $(c_{k_1'}^{(A)}, c_{k_1'}^{(B)})$ both varying in the neighborhood of (1,0) [See below for justification]. We utilize the Bloch sphere scheme and represent, for example, $(c_{k_1}^{(A)}, c_{k_1}^{(B)})$ by a point on the surface of the sphere, parameterized by

$$\begin{pmatrix} c_{k_1}^{(A)} \\ c_{k_1}^{(B)} \end{pmatrix} = \begin{pmatrix} \cos\dfrac{\theta_{k_1}}{2} \\ e^{i\varphi_1} \sin\dfrac{\theta_{k_1}}{2} \end{pmatrix}.$$

Here, $\theta_{k_1}$ is small, meaning that the point is near the north pole of the sphere. Evaluation in such a scheme yields $<\rho> \approx 1 - \dfrac{<|c_k^{(B)}|^2>}{4}$ and $<\mu> \approx 1 - \dfrac{<|c_k^{(B)}|^2>}{2}$ ($<|c_k^{(B)}|^2>$ = average occupation probability on site 'B', for states in the neighborhood of north pole).

We apply Eqn. (D-2) and estimate $<|c_k^{(B)}|^2>$ using Δ = 28meV in the quantum dot and electron energy E ~ 17 meV according to the calculation of one-carrier state energy shown in Figure 8. This yields $<|c_k^{(B)}|^2> \sim 0.19$, justifying our assumption of the states being near the north pole. Therefore, $<\mu> \approx 1 - \dfrac{<|c_k^{(B)}|^2>}{2} = 0.9$, meaning a reduction in the interaction strength, due to state mixing in the two-band model, in comparison to that in the reduced model ($\mu = 1$).

c) Last, the above discussion gives an estimate of about 10% error in the interaction matrix element, in the reduced model. Since the trion state is determined by both the kinetic effect (discussed in a)) and interaction effect (discussed in b)), an error of similar or less magnitude is expected for the trion state. Thus our numerical work based on the reduced Dirac model provides a reasonable study of trions.

**Figure Captions**

**Fig. 1** The energy band profile of a type-I graphene QD.

**Fig. 2** $E_{binding}$ vs. both the electron and the hole confinement potential strength ($V_{e(h)}^{(0)}$), in the case where the QD radius is 300A. Three barrier heights are considered, namely, 28meV, 42meV, and 56meV. The calculation is performed in the Schrodinger model with the Gaussian and exponential functions.

**Fig. 3** $E_{binding}$ vs. R (QD radius). R here changes from 150A to 450A. The calculation is performed in the Schrodinger model with the Gaussian, exponential, and Bessel functions.

**Fig. 4** $F_0(r_1, r_2, r_3 = 0)$ in the Schrodinger model, in the case where the QD radius is 300A. **a)** $F_0$ calculated using the Gaussian functions. **b)** $F_0$ calculated using the exponential functions.

**Fig. 5** $F_0(r_1, r_2, r_3 = 300A)$ in the Schrodinger model, in the case where the QD radius is 300A. **a)** $F_0$ calculated using the Gaussian functions. **b)** $F_0$ calculated using the exponential functions.

**Fig. 6** $F_0(r_1 = 0, r_2 = 0, r_3)$ in the Schrodinger model, in the case where the QD radius is 300A. $F_0$ is calculated using only the Gaussian functions.

**Fig. 7** $F_0(r_1 = 0, r_2 = 300A, r_3)$ in the Schrodinger model, in the case where the QD radius is 300A. $F_0$ is calculated using only the Gaussian functions.

**Fig. 8** $E_0^{(c)}$ (ground state energy of the confined electron) vs. R (QD radius), calculated in both the Schrodinger and Dirac models.

**Fig. 9** $\varphi_0^{(c)}$ (ground state wave function of the confined electron) calculated in both the Schrodinger and Dirac models, in the case where the QD radius is 150A.

**Fig. 10** $E_{binding}$ vs. R (QD radius). R here changes from 150A to 450A. The calculation is performed with the Bessel functions, in both the Schrodinger and Dirac models.

**Fig. 11** $F_0(r_1, r_2, r_3 = 0)$ calculated using the Bessel functions, in the case where the QD radius is 150A. **a)** $F_0(r_1, r_2, r_3 = 0)$ in the Schrodinger model. **b)** $F_0(r_1, r_2, r_3 = 0)$ in the Dirac model.

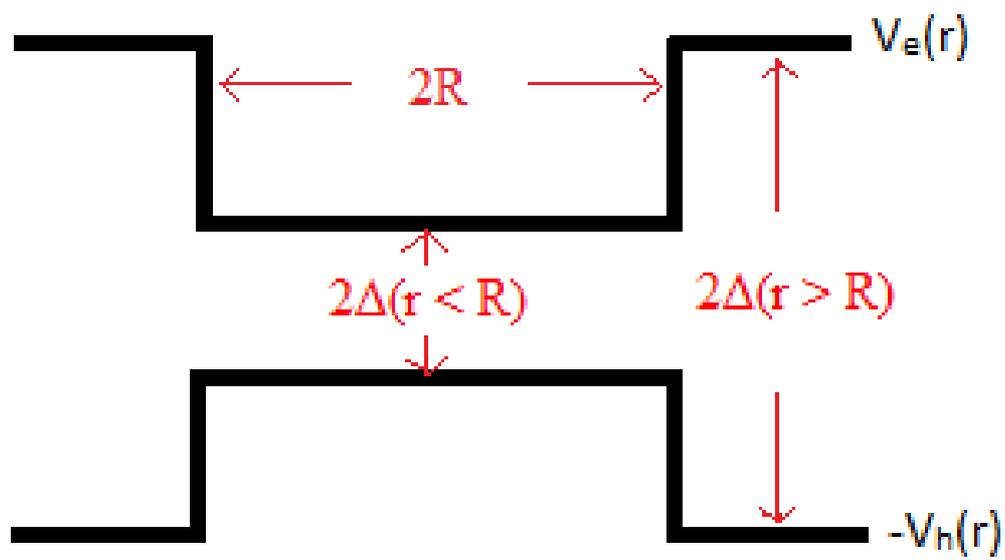

Figure 1

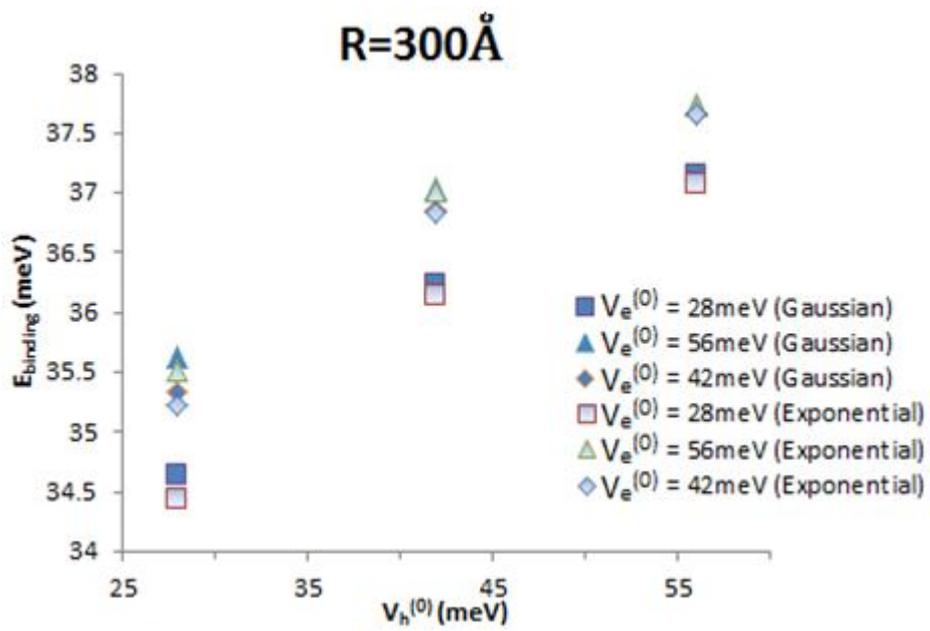

Figure 2

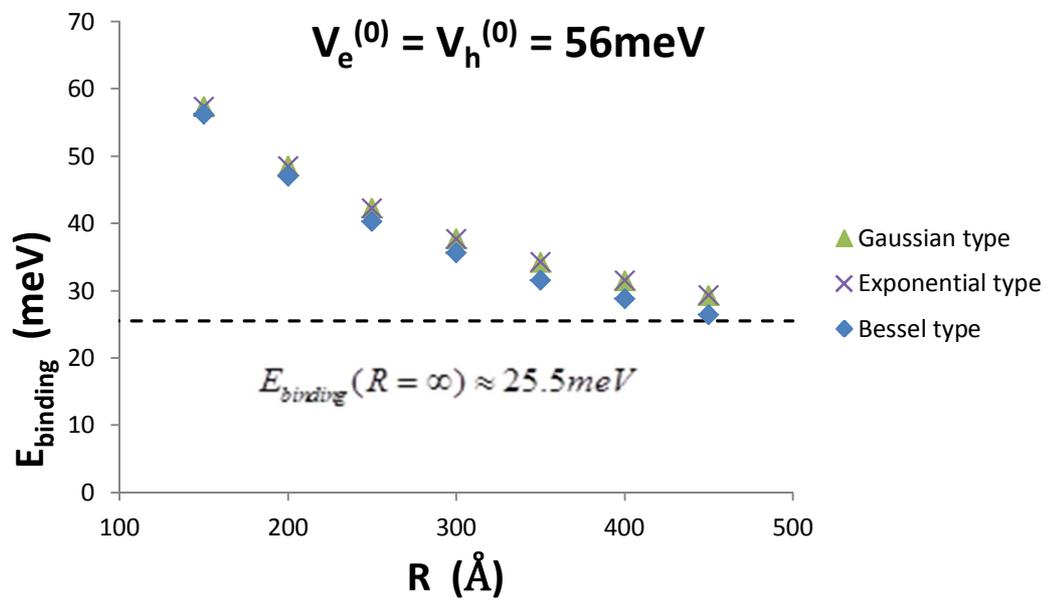

Figure 3

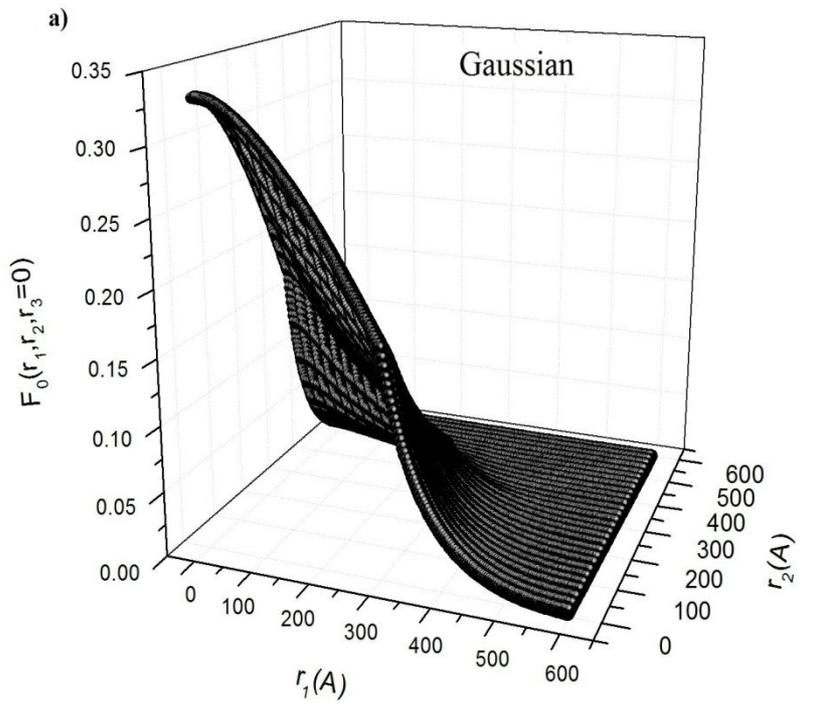

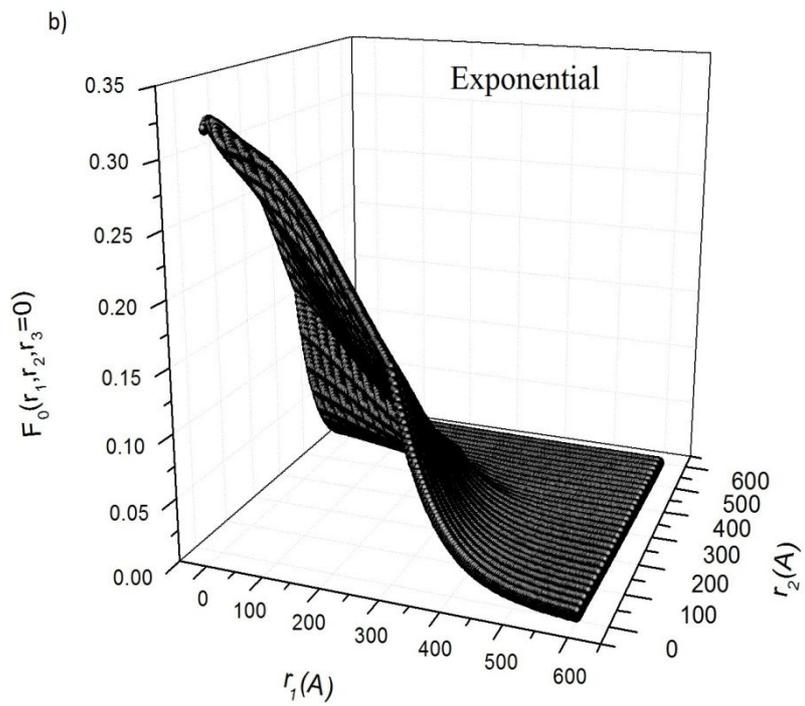

Figure 4

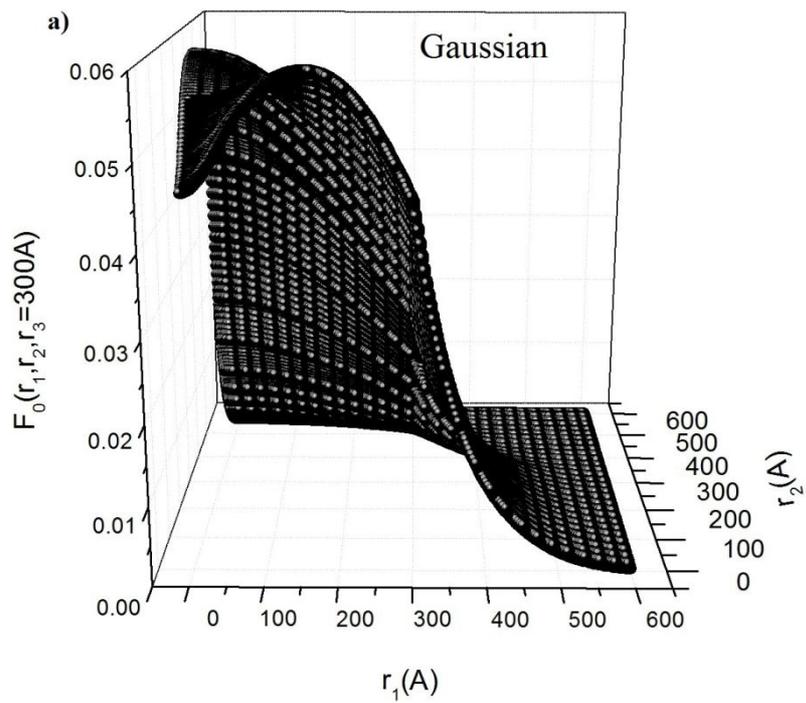

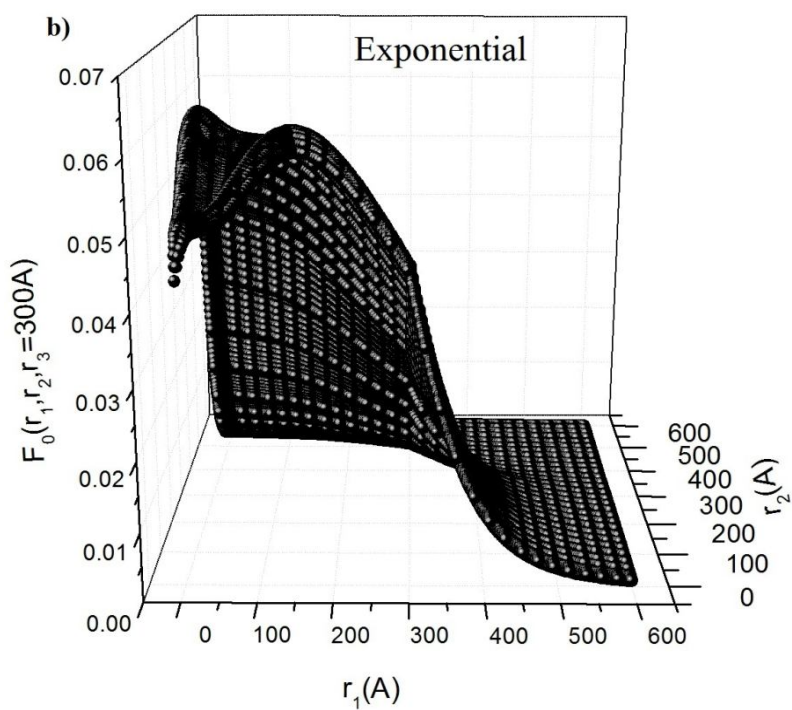

Figure 5

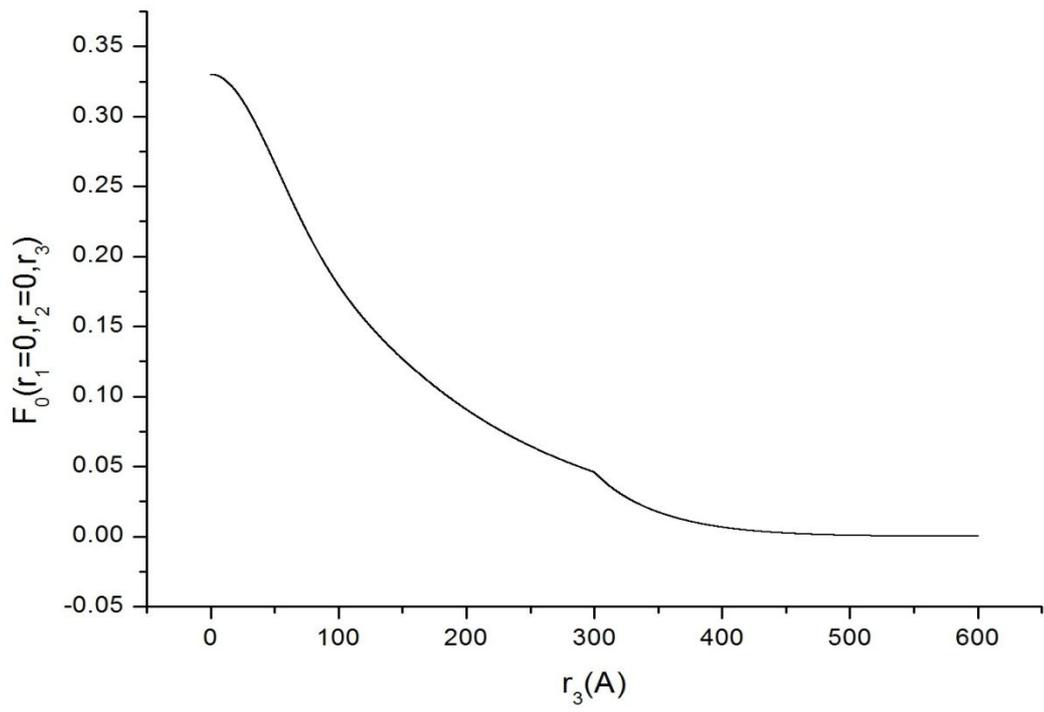

Figure 6

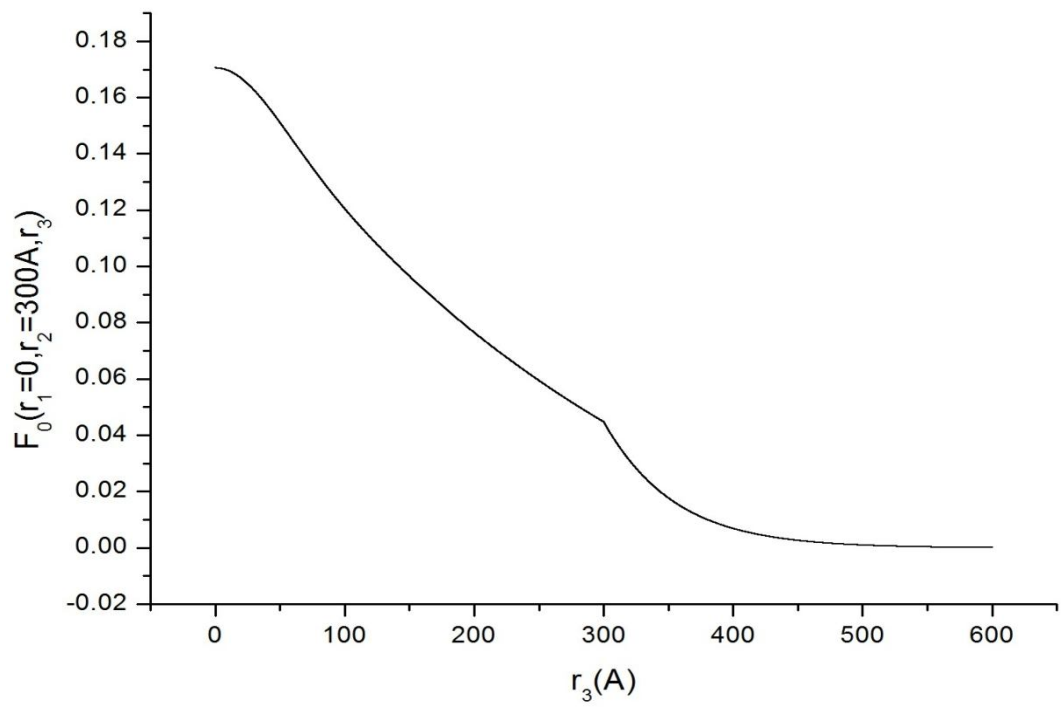

Figure 7

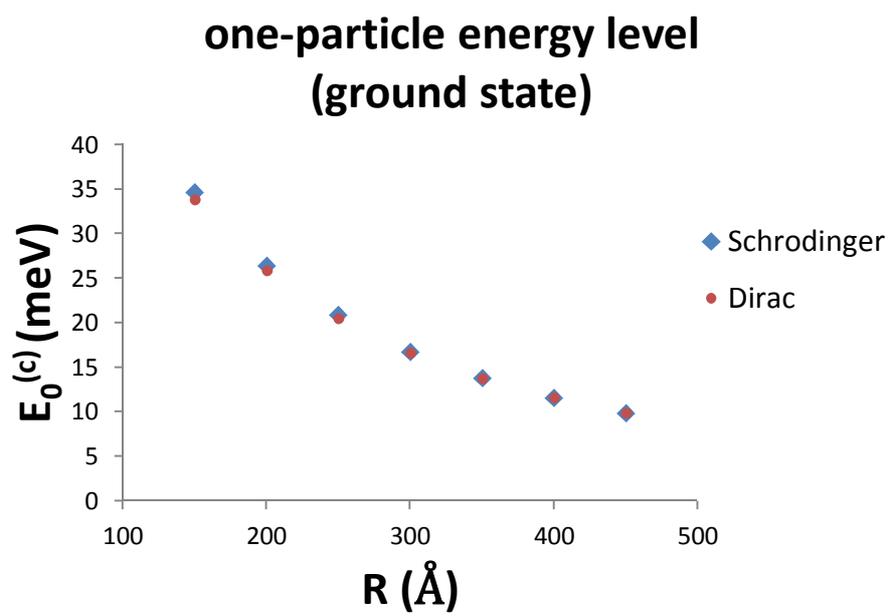

Figure 8

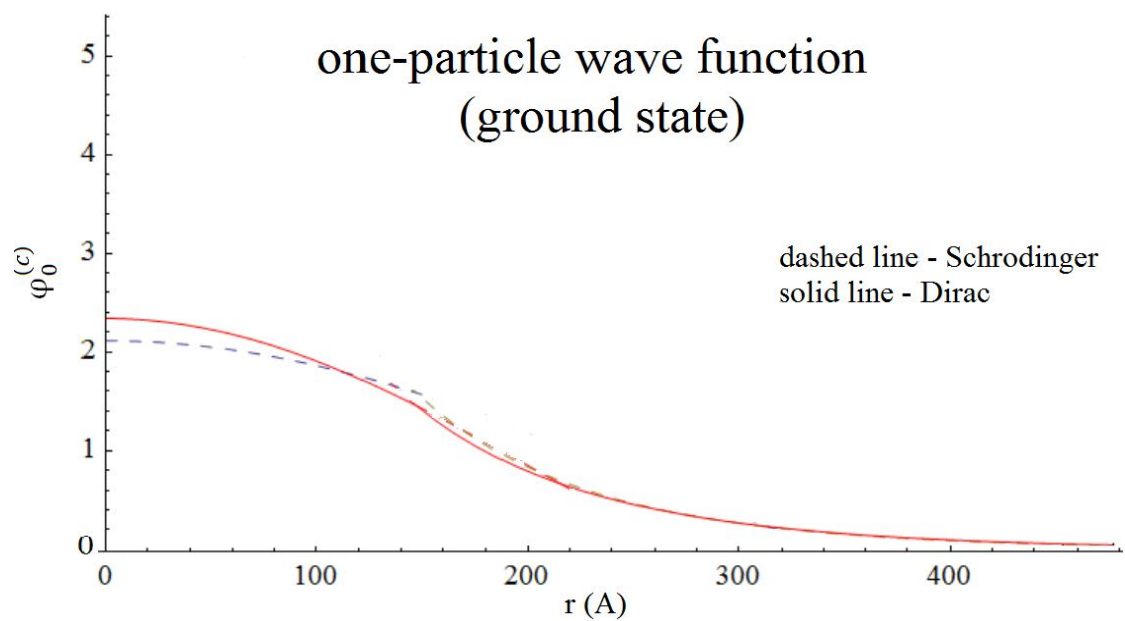

Figure 9

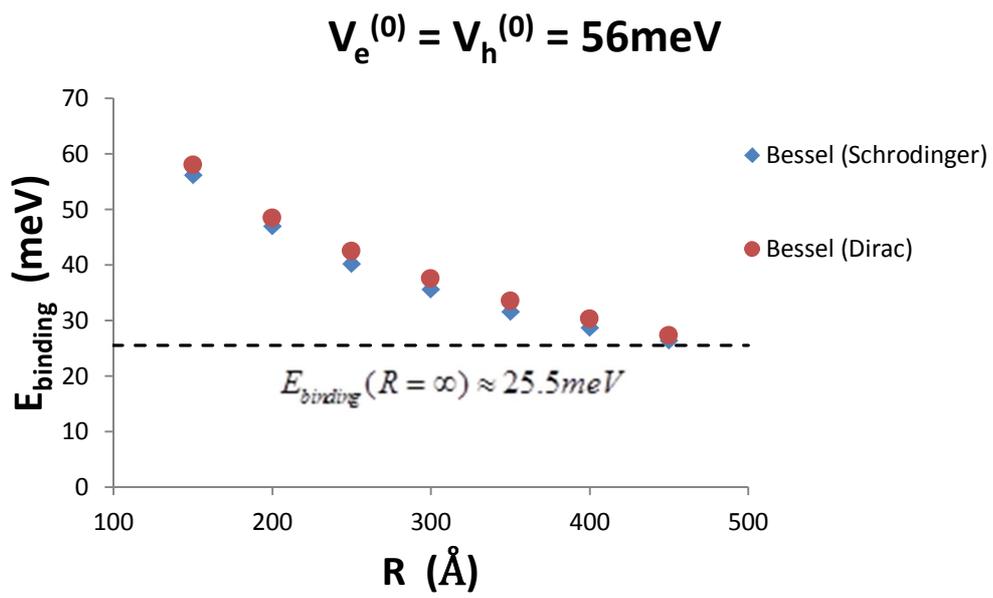

Figure 10

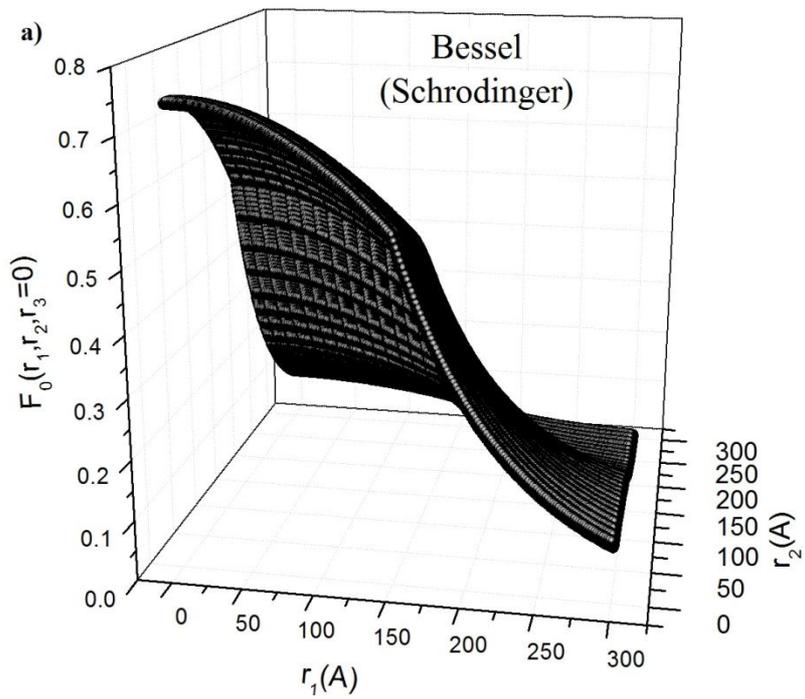

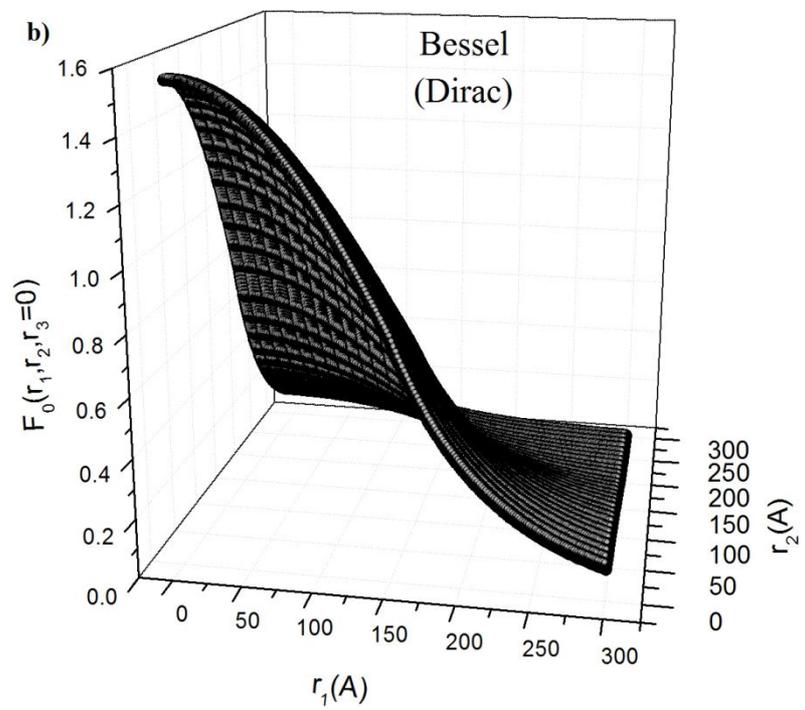

Figure 11